%% file: Paper_Klatt.tex
\documentclass[ASNA,twocolumn]{USG} 
\usepackage{anyfontsize} %

\usepackage{siunitx}
\usepackage{colortbl}   
\usepackage{xcolor}     

\usepackage{steinmetz}
\graphicspath{{./images/}}
\usepackage{graphicx}
\usepackage{subcaption}

\articletype{RESEARCH ARTICLE}%

\received{-}
\revised{-}
\accepted{-}
\journal{Submitted for Publication}
\volume{0}
\copyyear{2026}
\startpage{1}
\articledoi{0000}

\begin{document}
\title{Methodological Assessment of Instrumented Indentation for Spatially Resolved Mesoscale Stiffness and Creep in Short Fiber-Reinforced Injection-Molded PBT Components}

\transtitle{Manuscript Title}
\subtranstitle{trans-subtitle}
\author[1]{Eduard Klatt}[https://orcid.org/0000-0001-5690-017X]
\author[1]{Natalie Rauter}[https://orcid.org/0000-0003-1704-1426]

\authormark{KLATT \textsc{et al.}}
\titlemark{METHODOLOGICAL ASSESSMENT OF INSTRUMENTED INDENTATION FOR SPATIALLY RESOLVED MESOSCALE STIFFNESS AND CREEP IN SHORT FIBER-REINFORCED INJECTION-MOLDED PBT COMPONENTS}

\address[1]{\orgdiv{Helmut-Schmidt-University / University of the Federal Armed Forces Hamburg, }\orgname{Chair of Solid Mechanics, }%
\orgaddress{\state{Holstenhofweg 85, 22043 Hamburg, }\country{Germany}}}

\corres{Natalie Rauter  (\email{natalie.rauter@hsu-hh.de})}


\presentaddress{This is sample for present address text this is sample for present address text.}



\transkeywords{Instrumented Indentation | Mesoscale Testing | Fiber-reinforced Polymers | Spatially Resolved Viscoelasticity}

\abstract[ABSTRACT]{Instrumented indentation enables assessment of local mechanical properties in heterogeneous polymer systems, but specimen preparation and the selected indentation protocol significantly influence the quantitative interpretation of the measured response. This study develops and evaluates an experimental methodology for spatially resolved mesoscale characterization of both unreinforced and short fiber-reinforced composites. An unreinforced material system is used as a reference material to systematically examine the effects of indentation spacing, surface preparation, maximum indentation load, temporal loading parameters, and unloading evaluation. Indentation loads of $\SI{500}{\milli\newton}$, $\SI{1000}{\milli\newton}$, and $\SI{1750}{\milli\newton}$ are investigated in complementary experimental campaigns. The apparent indentation modulus exhibits moderate sensitivity to indentation load and loading time within the studied ranges. In contrast, indentation depth and the creep-related depth increment display greater sensitivity to the applied load history. Surface polishing minimizes topographical disturbances and enables reliable identification of residual indentation profiles and pile-up. Based on these methodological findings, a final load–hold–unload protocol with an indentation load of $\SI{1750}{\milli\newton}$, a loading time of $\SI{6}{\second}$, a holding period of $\SI{300}{\second}$, and an unloading time of $\SI{4}{\second}$ is applied to both unreinforced material and short fiber-reinforced composites. The mean apparent indentation modulus increases from $\SI{2.91}{\giga\pascal}$ for unreinforced material to $\SI{6.65}{\giga\pascal}$ for short fiber-reinforced composites, while the mean depth of indentation creep decreases from $\SI{2.36}{\micro\metre}$ to $\SI{1.91}{\micro\metre}$. The coefficient of variation also increases for both quantities, indicating that short-fiber reinforcement enhances mean local stiffness and introduces pronounced spatial heterogeneity in both stiffness-related and time-dependent mesoscale indentation responses.}





\copyright{Copyright Statement
\\[5pt]
  ©  2024 XXXX}


\maketitle


\section{Introduction}\label{sec:introduction}

Short-fiber-reinforced composites (SFRCs) are widely used in mechanically loaded components because they combine low density, efficient processability, and relatively high stiffness and strength. Injection-molded SFRCs, in particular, present considerable potential for lightweight structural applications with complex geometries. This study examines two injection-molded polybutylene terephthalate (PBT) material systems: unreinforced PBT (PBT~GF0) and short-glass-fiber-reinforced PBT with a nominal glass-fiber mass fraction of $30~\mathrm{wt.\%}$ (PBT~GF30). Both systems utilize the same thermoplastic matrix, which allows for a direct comparison of the effects of short-glass-fiber reinforcement on the local mechanical response at the mesoscale.

The mechanical behavior of PBT~GF30 is influenced not only by the properties of the polymer matrix and reinforcement but also by the heterogeneous microstructure formed during injection molding. Variations in fiber content, fiber orientation, matrix morphology, and fiber--matrix interactions produce spatially heterogeneous mechanical properties. This heterogeneity is especially pronounced in the characteristic skin--core morphology of injection-molded SFRCs, where flow-induced shear aligns fibers predominantly along the melt-flow direction in the outer layers, while more transverse or less strongly aligned fiber orientations are observed in the core~\cite{Advani1987, Banik2005}.

This heterogeneity becomes particularly important when the polymer matrix exhibits time-dependent behavior. Thermoplastic matrices undergo viscoelastic deformation under sustained mechanical loading, while the incorporation of stiff fibers modifies both the overall stiffness and the resulting time-dependent deformation. Consequently, characterization based only on spatially averaged component-level properties does not adequately capture the local material response required for micromechanical analysis or spatially resolved material modeling~\cite{Kaliske2000}.

Instrumented indentation is an effective technique for investigating mechanical behavior at the mesoscale. During these experiments, the continuously recorded force–displacement response allows for the determination of local stiffness-related properties from the unloading segment and facilitates the assessment of time-dependent deformation by analyzing displacement evolution under sustained load. The Oliver–Pharr methodology is commonly employed to evaluate indentation modulus and hardness from indentation measurements~\cite{Oliver1992, Oliver2004}. Furthermore, extended holding segments at maximum load enable quantification of the time-dependent response of SFRC through indentation creep measurements~\cite{Tranchida2007}.

The indentation response of polymers is highly sensitive to experimental and evaluation parameters~\cite{Hardiman2016}, e.g. the quantitative indentation response of polymers is sensitive to specimen surface condition, maximum load, and testing speed~\cite{Huszar2024}. Surface roughness affects initial contact detection and local contact geometry. Insufficient spacing between adjacent indentations can lead to interactions with previously plastically deformed regions. Additionally, the loading rate, holding period, unloading duration, and the chosen unloading-curve range influence the evaluated indentation modulus, as time-dependent deformation may persist throughout the entire indentation cycle~\cite{Hardiman2016, Hochstetter1999}. Hence, the prescribed load-hold-unload history and the unloading rate can significantly influence the separation of elastic and time-dependent deformation~\cite{Jakes2024, Sarkar2024, SarkarVerma2026}. Direct observations of viscoelastoplastic polymers also indicate that sink-in and pile-up phenomena can coexist during the indentation cycle, resulting in contact geometries that may deviate from the assumptions of conventional Oliver-Pharr evaluation~\cite{Moharrami2014,SarkarPantNanavati2024, Tranchida2007}.

Spatial indentation is a well-established method for characterizing heterogeneous polymers and polymer-matrix composites~\cite{GaillardAmiot2020, Grondin2023}. For instance, Rauter~\cite{Rauter2021} demonstrated spatially resolved indentation modulus mapping of injection-molded SFRC on the mesoscale and quantified the effective material region represented by the indentation response. Furthermore, indentation creep has been applied to SFRC~\cite{Shen2022} and has also been reported for polymers filled with glass fiber in studies of radiation-induced material modification~\cite{Ovsik2016}.

These methodological factors are particularly critical when interpreting spatial variations in indentation properties as indicators of intrinsic material heterogeneity~\cite{Klatt2025a, Rauter_habil2023}. It is essential to distinguish experimental scatter resulting from surface preparation or inappropriate indentation parameters from the true spatial variability of the material under investigation~\cite{Hardiman2016}. Therefore, a consistent indentation protocol should be established prior to statistical evaluation of differences between unreinforced material and SFRC.

The current literature lacks an integrated experimental methodology that systematically assesses the influence of surface preparation, indentation spacing, maximum load, temporal test parameters, and unloading evaluation, followed by the application of a unified protocol to spatially map both the apparent indentation modulus and the long-hold indentation creep response of matched unreinforced material and SFRC. This study addresses this gap by combining a systematic methodological assessment with a spatially resolved application to injection-molded PBT. Unreinforced PBT~GF0 is used as the reference system to evaluate indentation spacing, surface preparation, indentation load, temporal test parameters, and unloading evaluation. Based on these measurements, a final indentation protocol is developed and applied to both PBT~GF0 and glass fiber reinforced PBT~GF30.

Subsequent analyses focus on the apparent indentation modulus $E_\mathrm{ind}$ and the absolute indentation creep depth increment $\Delta h_\mathrm{CR}$. Here, spatially distributed measurements on polished specimen cross-sections are conducted to determine mean values, standard deviations, coefficients of variation (CoV), and spatial distributions of $E_\mathrm{ind}$ and $\Delta h_\mathrm{CR}$. This approach enables a quantitative assessment of the influence of short-fiber reinforcement on both the average local response and its spatial variability. Additionally, the spatially resolved measurements yield statistical descriptors and spatial response maps of local material behavior, providing an experimental foundation for subsequent stochastic analyses that explicitly consider spatial material variability \cite{OstojaStarzewski2006}.

Accordingly, the remainder of this paper is organized as follows. Section~\ref{sec:framework} provides the main theoretical aspects. All information about the investigated material systems, specimen preparation, and the used laboratory testing equipment are introduced in Section~\ref{sec:methodology}. Section~\ref{sec:methodological_assessment} presents and discusses the methodological assessment performed on PBT~GF0 and consolidates their implications into the final indentation protocol. The spatially resolved results for PBT~GF0 and PBT~GF30 are presented in Section~\ref{sec:application} including the discussion of the material-related findings and its limitations. This is followed by the main conclusions in Section~\ref{sec:conclusions}.

\section{Theoretical framework of mesoscale instrumented indentation testing}\label{sec:framework}

\subsection{Scale separation and mesoscale indentation concept}
\label{subsec:scale_separation}

Following the multi-scale concept, the mechanical response of SFRC is influenced by the spatial scale at which the heterogeneous microstructure is examined. At the microscale, individual constituents and their interfaces are explicitly resolved. In contrast, structural testing at the component level yields an effective response averaged over a larger representative material volume. Between these two extremes, a mesoscale description characterizes the apparent local response of a heterogeneous material region without resolving each constituent individually~\cite{Constantinides2006, OstojaStarzewski2006, RauterIndentation2021, Ulm2010}. 

\begin{figure}[t]
   \centering
   \def\svgwidth{0.99\columnwidth}
    \graphicspath{{images/}}
    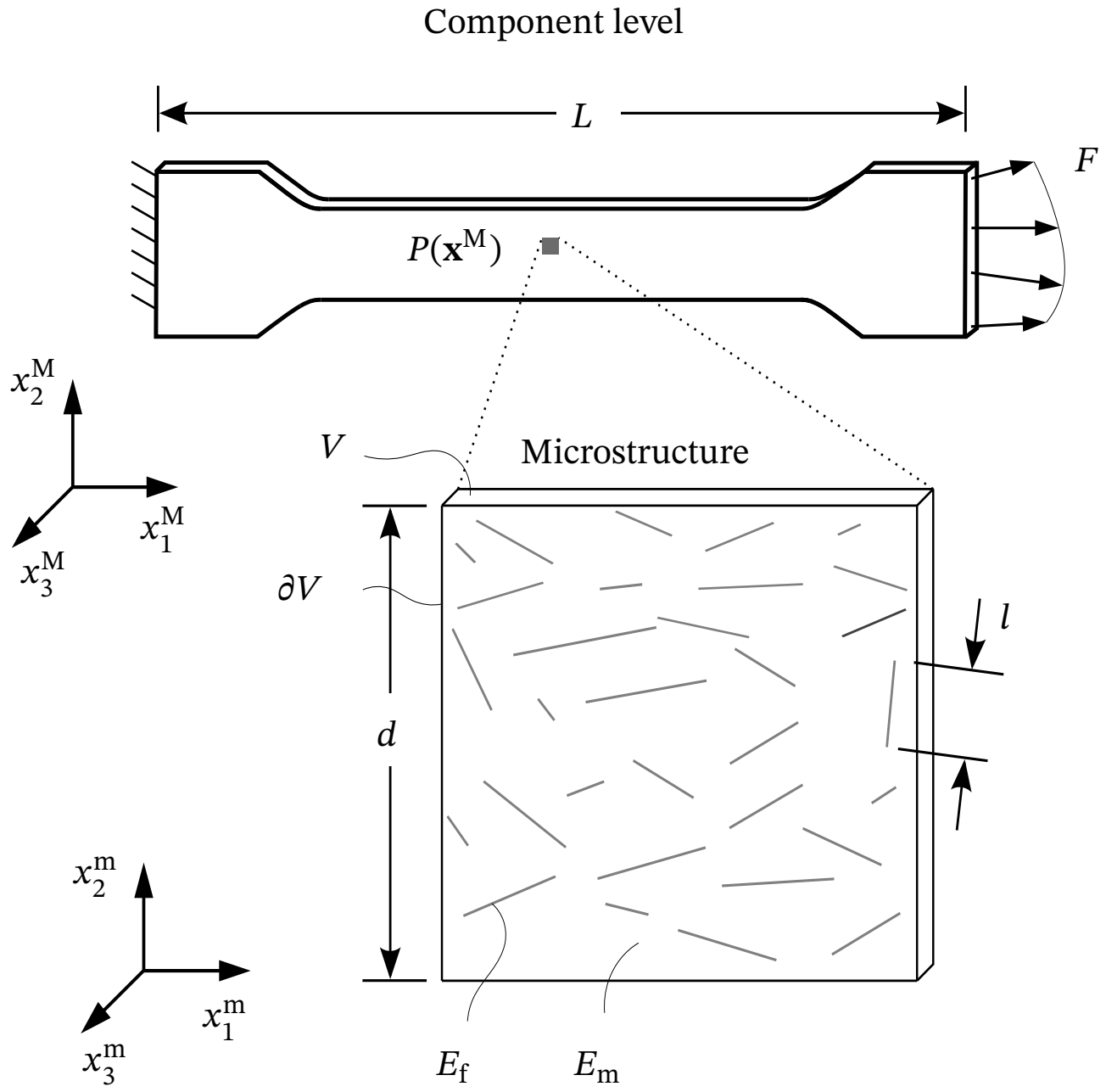
    \caption{A schematic illustration demonstrates scale separation in a short-fiber-reinforced composite. The macroscopic component is defined by the length scale $L$, while the local material region $V$, bounded by $\partial V$ and characterized by dimension $d$, contains glass fibers embedded within the polymer matrix. The characteristic microstructural length $l$ represents the scale of the underlying heterogeneities. The elastic moduli of the fibers and matrix are denoted as $E_\mathrm{f}$ and $E_\mathrm{m}$, respectively. The coordinate systems $\mathbf{x}^{\mathrm{M}}$ and $\mathbf{x}^{\mathrm{m}}$ correspond to the component level and local material descriptions, respectively.}
    \label{fig:scale_of_separation}
\end{figure}

For a heterogeneous material with characteristic microstructural length $l$, mesoscale length $d$, and characteristic component dimension $L$, the classical separation-of-scales condition is expressed as \cite{huet1990application,zohdi2008introduction}
\begin{equation}
    l \ll d \ll L ~.
    \label{eq:separation_of_scales}
\end{equation}
Figure~\ref {fig:scale_of_separation} schematically illustrates the corresponding scale hierarchy and the relationship between the macroscopic component and the local heterogeneous material region.

In this context, $l$ denotes the characteristic length of the microstructural heterogeneity, including the glass fibers and their spatial arrangement, whereas $L$ represents the macroscopic specimen or component-level dimension. Under the assumed separation of scales, the intermediate length $d$ characterizes the mesoscale, at which the apparent local mechanical response reflects the collective influence of the heterogeneous microstructure rather than the properties of individual constituents~\cite{RauterIndentation2021}.

The characteristic microstructural length scale $l$ differs fundamentally between the two material systems examined in this study. In unreinforced PBT (PBT~GF0), $l$ is determined by the polymer matrix morphology, including semi-crystalline domain and spherulite sizes as well as the arrangement of crystalline lamellae, and typically falls within the sub-micron to low-micrometer range~\cite{Grondin2023, Jakes2024}. For a given projected indentation area, the resulting scale ratio $d/l$ is therefore very large, so the indentation process samples an effectively homogeneous material continuum. In contrast, in PBT~GF30, the characteristic length $l$ is defined by the mean diameter of the glass fibers ($\bar{d} \approx \SI{10}{\micro\meter}$) and their spatial distribution. Under identical indentation loads and projected contact areas, the indentation-affected volume in PBT~GF30 contains only a few fibers and adjacent matrix regions. Thus, the $d/l$ ratio is much smaller for PBT~GF30, rendering the mesoscale response highly sensitive to local microstructural variations and contributing to the observed experimental parameter scatter.

Instrumented indentation is applied within the mesoscale regime in this study. The volume of material affected by each indentation comprises contributions from the polymer matrix and, for PBT~GF30, from locally varying fiber configurations that constrain the expanding elastic-plastic deformation fields~\cite{GaillardAmiot2020, Slagter2023}. Consequently, the measured response is interpreted not as a phase-specific constituent property but as an apparent local response associated with the indentation-affected material region~\cite{Constantinides2006, RauterIndentation2021}. Spatial repetition of this measurement across the specimen cross-section provides a mesoscale representation of local mechanical variability while preserving positional context within the macroscopic specimen.

\subsection{Indentation modulus and indentation creep}\label{subsec:indentation_quantities}

Instrumented indentation measurements were evaluated using the recorded indentation force $P$ and indentation depth $h$. All experiments employed a Berkovich diamond indenter. The apparent indentation modulus $E_\mathrm{ind}$ was determined from the unloading phase of the force--displacement response using the Oliver–Pharr approach \cite{Oliver1992, Oliver2004}.

\begin{figure}[t]
   \centering
   \def\svgwidth{0.95\columnwidth}
    \graphicspath{{images/}}
    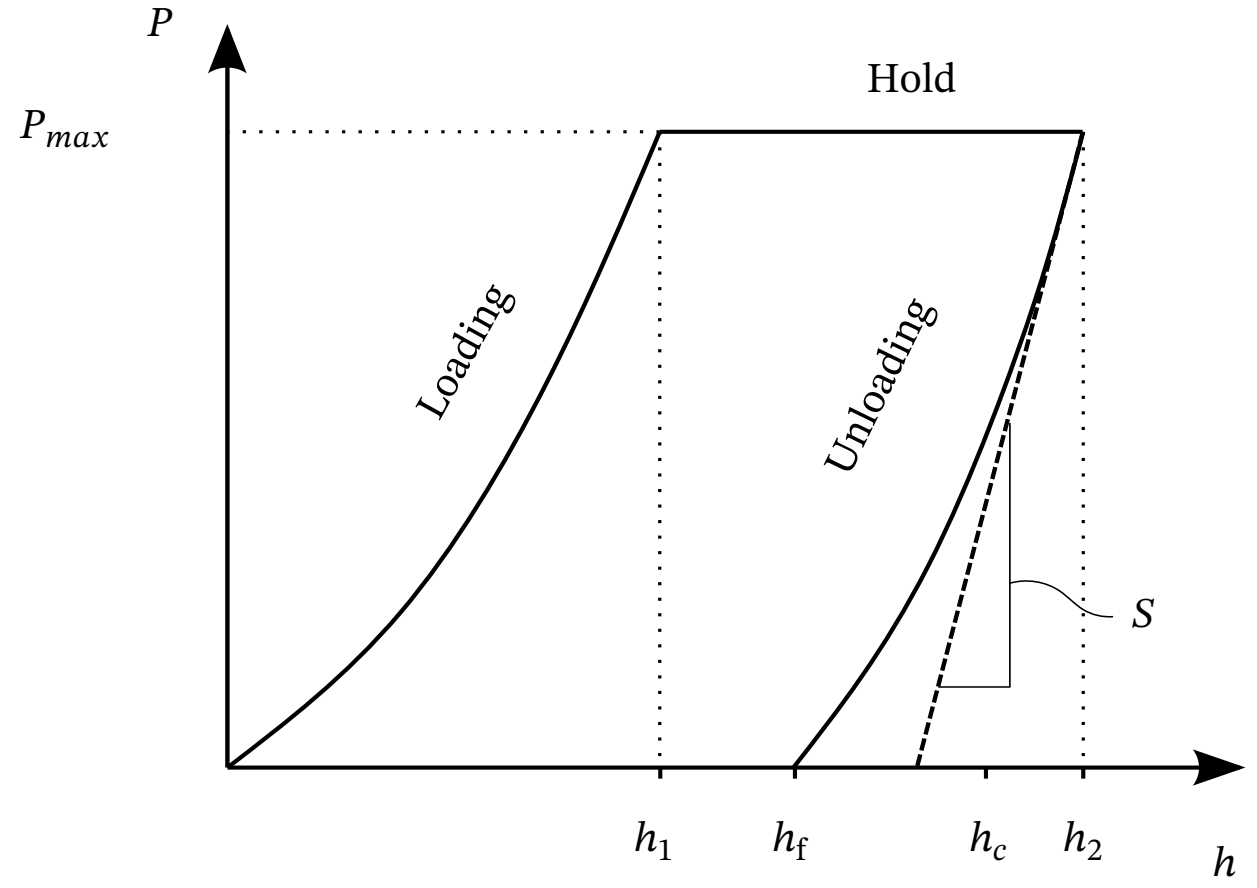
    \caption{Schematic force--displacement response during instrumented indentation illustrates the loading, holding, and unloading segments. During the holding period at $ P_\mathrm {max}$, the indentation depth increases from $h_1$ to $h_2 = h_\mathrm{max}$, which defines the indentation creep depth increment $\Delta h_\mathrm{CR} = h_2 - h_1$. The contact stiffness $S$ is determined from the initial unloading slope.  The parameter $h_\mathrm{f}$ indicates the extrapolated residual depth, while $h_\mathrm{c}$ refers to the calculated contact depth as described in~\cite{Oliver1992, Oliver2004}.}
    \label{fig:indentation_load_displacement}
\end{figure}

Figure~\ref{fig:indentation_load_displacement} schematically illustrates the characteristic force--displacement response during a load--hold--unload indentation cycle. The holding segment at constant maximum load $P_\mathrm{max}$ quantifies the additional time-dependent indentation displacement, while the subsequent unloading response provides the contact stiffness $S$ needed to determine the apparent indentation modulus $E_\mathrm{ind}$.

 The unloading curve is approximated using a power-law relation of the form
\begin{equation} 
P(h) = \alpha \left(h-h_\mathrm{f}\right)^m~, 
\label{eq:unloading_power_law} 
\end{equation}
where $\alpha$ and $m$ are fitting parameters, and $h_\mathrm{f}$ denotes the residual indentation depth obtained by extrapolating the unloading curve. The contact stiffness $S$ is subsequently evaluated from the derivative of Equation~\eqref{eq:unloading_power_law} at the maximum indentation depth,
\begin{equation} 
S = \left. \frac{\mathrm{d}P}{\mathrm{d}h} \right|_{h=h_\mathrm{max}}~. 
\label{eq:contact_stiffness} 
\end{equation}

\begin{figure}[h]
   \centering
   \def\svgwidth{0.80\columnwidth}
    \graphicspath{{images/}}
    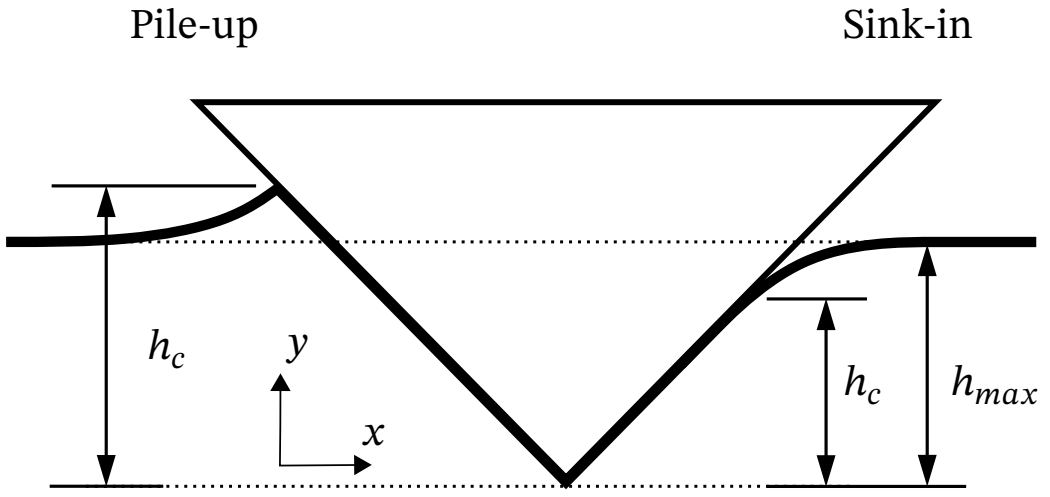
    \caption{Schematic comparison of pile-up and sink-in during instrumented indentation. Pile-up produces upward displacement of material around the indenter, potentially increasing the actual geometrical contact depth and projected contact area beyond the values estimated using the conventional Oliver--Pharr procedure. In contrast, sink-in produces a downward displacement of the surrounding surface, which is accounted for by the conventional contact-depth approximation under the assumed elastic contact conditions~\cite{Oliver1992, Oliver2004}.}
    \label{fig:sinkin_pileup_scheme}
\end{figure}

Based on Sneddon's analytical solution for the boundary deflection of an elastic half-space \cite{Sneddon1965}, the corresponding contact depth is calculated as
\begin{equation} 
h_\mathrm{c} = h_\mathrm{max} - \varepsilon \,\frac{P_\mathrm{max}}{S}~.
\label{eq:contact_depth} 
\end{equation}
Here, $P_\mathrm{max}$ denotes the maximum indentation load, and $\varepsilon=0.75$ is the contact-depth correction coefficient adopted for the Berkovich indentation evaluation~\cite{Jakes2018, Oliver1992, Tranchida2007}. Figure~\ref{fig:sinkin_pileup_scheme} schematically illustrates the influence of sink-in and pile-up on the contact geometry. The conventional Oliver--Pharr contact-depth approximation accounts for sink-in; however, pile-up can result in the actual contact depth and projected contact area exceeding their estimated values.

The projected contact area $A_\mathrm{p}(h_\mathrm{c})$ is determined using the calibrated area function of the Berkovich indenter, which accounts for deviations from the ideal pyramidal geometry, including the finite tip radius~\cite{Oliver2004}. Because the estimated contact depth may be inaccurate in the presence of pile-up, the resulting indentation modulus is interpreted as an apparent material property at the mesoscale. The associated limitations are discussed in Section~\ref{subsec:evaluation_limitations}.

Assuming an isotropic contact relation, the reduced indentation modulus is given by 
\begin{equation} 
E_\mathrm{r} = \frac{\sqrt{\pi}}{2\,\beta} \frac{S} {\sqrt{A_\mathrm{p}(h_\mathrm{c})}}~.
\label{eq:reduced_indentation_modulus} 
\end{equation} 
In this context, a symmetry correction factor of $\beta = 1$ is incorporated. This assumption is widely accepted for Berkovich geometries~\cite{Hay1999, RauterIndentation2021, Troyon2006}.

The indentation modulus $E_\mathrm{ind}$ of the specimen is calculated by accounting for the elastic properties of the diamond indenter and the reduced indentation modulus, as expressed by
\begin{equation} 
\frac{1}{E_\mathrm{r}} = \frac{1-\nu^2}{E_\mathrm{ind}} + \frac{1-\nu_\mathrm{i}^2}{E_\mathrm{i}}~.
\label{eq:indentation_modulus} 
\end{equation} 
Here, $E_\mathrm{i}=\SI{1141}{\giga\pascal}$ and $\nu_\mathrm{i}=\SI{0.07}{}$ denote the Young's modulus and Poisson's ratio of the indenter~\cite{Oliver1992}, respectively, while $\nu$ refers to the assumed Poisson's ratio of the specimen. For PBT~GF0, the Poisson's ratio is $\nu_{\text{PBT~GF0}}=\SI{0.41}{}$~\cite{Rauter_habil2023}, and for PBT~GF30, $\nu_{\text{PBT~GF30}}=\SI{0.38}{}$~\cite{Rauter2021}.

In addition to the stiffness-related response, the time-dependent indentation behavior is evaluated using the holding segment at $P=P_\mathrm{max}$ to address the pronounced viscoelasticity of the polymer matrix \cite{Hardiman2016, Slouf2023, Tranchida2007}. The indentation depth at the beginning of the holding period is denoted as $h_1$, while $h_2=h_\mathrm{max}$ represents the maximum indentation depth reached at the end of the holding period. The increment in absolute indentation creep depth is defined as
\begin{equation} 
\Delta h_\mathrm{CR} = h_2-h_1~.
\label{eq:absolute_indentation_creep} 
\end{equation} 
This value quantifies the additional indentation depth accumulated during the holding period and is used as the primary measure of the time-dependent indentation response in this study. To account for the indentation depth already reached at the beginning of the holding period, the relative indentation creep $C_\mathrm{IT}$ is additionally defined as~\cite{Slouf2023}
\begin{equation}
C_\mathrm{IT} = \frac{h_2-h_1}{h_1}\cdot 100\,\% = \frac{\Delta h_\mathrm{CR}}{h_1}\cdot 100\,\%~.
\label{eq:relative_indentation_creep}
\end{equation}
Here, $h_1$ denotes the indentation depth at the beginning of the holding period. The relative indentation creep $C_\mathrm{IT}$ describes the percentage increase in indentation depth during the constant-load holding period. In contrast to $\Delta h_\mathrm{CR}$, which is expressed in units of length, $C_\mathrm{IT}$ provides a dimensionless measure of the time-dependent indentation response.

\subsection{Evaluation limitations associated with pile-up and viscoelastic unloading}
\label{subsec:evaluation_limitations}

The Oliver--Pharr evaluation method assumes that the projected contact area is determined from the contact depth and the elastic unloading response. Although this approach accounts for elastic sink-in around the indenter, it becomes inaccurate when material is displaced upward around the residual indentation imprint, resulting in pile-up \cite{Hardiman2017, Oliver1992, Oliver2004, Tranchida2007}. In these cases, the actual projected contact area can exceed the area estimated from $h_\mathrm{c}$, leading to an underestimation of the contact area and a corresponding overestimation of the derived modulus and hardness \cite{Bolshakov1998, Moharrami2014, Pharr2002}.

This limitation is particularly relevant for polymers and polymer-matrix composites, where plastic and time-dependent deformation may persist throughout the indentation cycle~\cite{Hardiman2016, Tranchida2007}. In SFRC, the heterogeneous fiber--matrix microstructure may additionally modify the local deformation field and the resulting surface displacement around the indenter~\cite{Hardiman2016, Hardiman2017}. Within this work, the residual surface geometry was not used to correct the projected contact area, and no pointwise contact-area correction was performed for the spatial indentation measurements. Consequently, the resulting modulus is consistently interpreted as an apparent indentation modulus $E_\mathrm{ind}$.

A second limitation arises from ongoing time-dependent deformation during unloading. If the viscoelastic displacement rate remains significant at the transition from the holding segment to unloading, the initial curvature of the force--displacement response may deviate from the predominantly elastic unloading response assumed in conventional evaluation. In severe cases, this persistent creep deformation can result in a negative initial unloading slope, commonly referred to as the viscoelastic "nose" effect, which precludes standardized stiffness determination \cite{Jakes2024, Tranchida2007}. Consequently, the choice of the fitting interval in the unloading phase can substantially affect the calculated contact stiffness $S$ and, therefore, $E_\mathrm{ind}$. 

\subsection{Statistical measures} \label{subsec:statistical_evaluation}

Spatial indentation data were analyzed using descriptive statistical measures to differentiate changes in mean local response from variations in spatial variability. For a quantity $X$ measured at $n$ valid indentation positions, the arithmetic mean is defined by
\begin{equation} 
\overline{X} = \frac{1}{n} \sum_{i=1}^{n} X_i~. 
\label{eq:mean} 
\end{equation}

The sample standard deviation is calculated as
\begin{equation} 
s_X = \sqrt{ \frac{1}{n-1} \sum_{i=1}^{n} \left(X_i-\overline{X}\right)^2 }~,
\label{eq:standard_deviation} 
\end{equation}

and the CoV is obtained from
\begin{equation} 
\mathrm{CoV}_X = \frac{s_X}{\overline{X}} \cdot 100\,\%~. 
\label{eq:cov} 
\end{equation}
The CoV facilitates a scale-independent comparison of relative spatial scatter between PBT~GF0 and PBT~GF30, as the mean values of both $E_\mathrm{ind}$ and $\Delta h_\mathrm{CR}$ differ substantially between these material systems due to the stiffening effect of the short-fiber reinforcement \cite{Fahrmeir2024}.

\section{Experimental setup}\label{sec:methodology}

\subsection{Material systems and specimen definition}\label{subsec:materials}

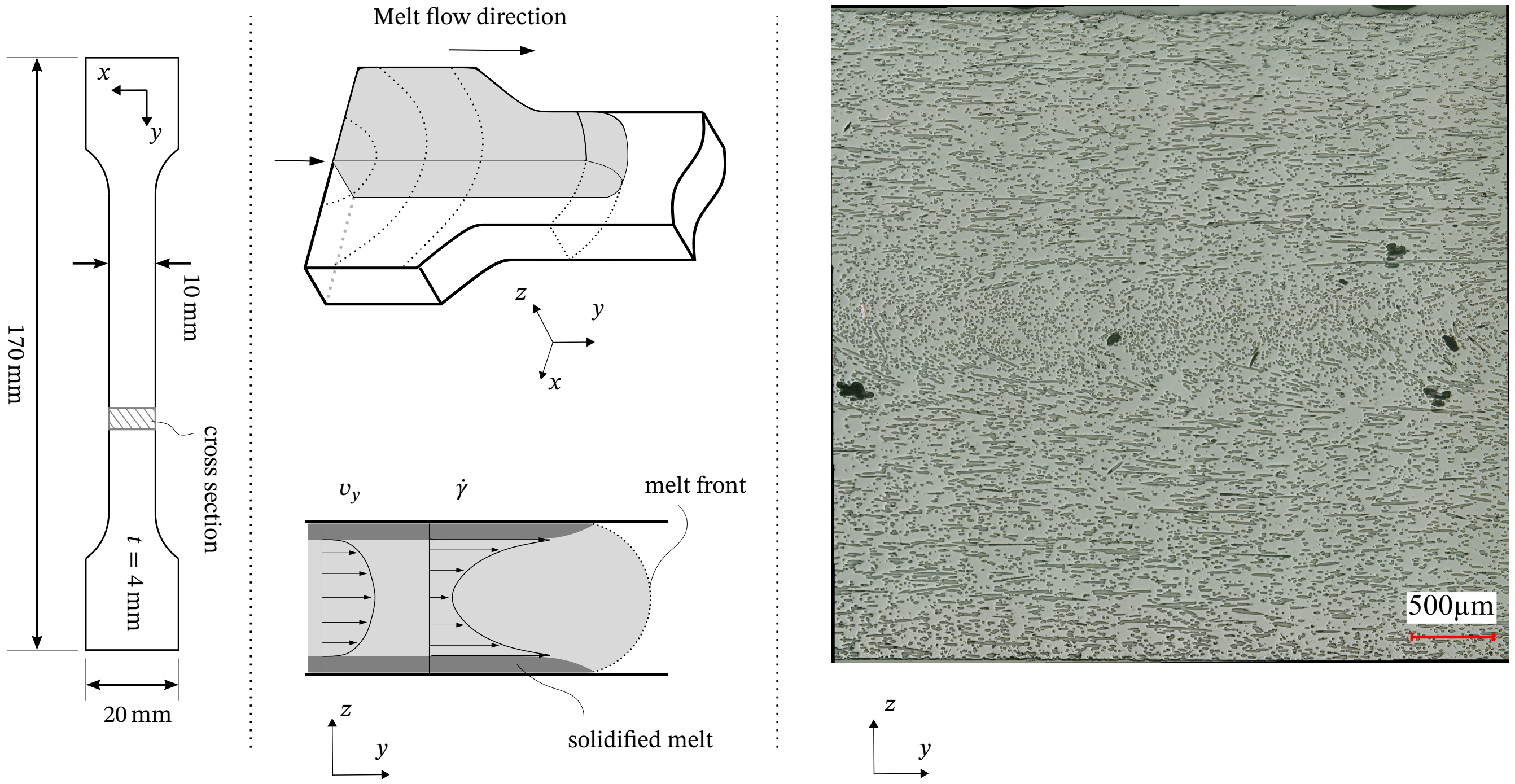
\begin{figure*}[t]
   \centering
    \def\svgwidth{0.99\textwidth}
    \graphicspath{{images/}}
    \input{images/03_Zugprobe_1A_plane_yz_mould-injection_6_paper.pdf_tex}
    \caption{Specimen geometry, injection-molding flow conditions, and microstructural heterogeneity of PBT~GF30. Left: Geometry of the standardized injection-molded tensile specimens, indicating nominal dimensions and the designated cross-sectional sampling location as specified by DIN EN ISO 527. Center: Schematic diagrams illustrating the principal melt-flow direction (upper) and the velocity distribution with the fountain-flow mechanism at the advancing melt front (lower). Right: Representative optical micrograph of PBT~GF30 in the longitudinal $y$--$z$ plane, highlighting the heterogeneous fiber arrangement resulting from injection molding.}
    \label{fig:injection_molding_flow}
\end{figure*}

The material system investigated in this study consists of a PBT matrix based on Ultradur B~4520 and short E-glass fibers as the reinforcement phase. This material system was used to produce two sets of injection-molded standardized tensile specimens conforming to DIN EN ISO 527-2 type 1A at Kunststoff-Zentrum Leipzig gGmbH using an Arburg Allrounder 370E injection molding machine. The injection-molding parameters are detailed in~\cite{Klatt2025}. Both PBT~GF0, designating the unreinforced specimens, and PBT~GF30, a reinforced material system with a nominal short-fiber mass fraction of $30~\mathrm{wt.\%}$, specimens were manufactured with identical nominal geometries, as depicted in Figure~\ref{fig:injection_molding_flow}, and comparable processing conditions to provide a consistent experimental basis for the subsequent indentation investigations.

PBT~GF0 serves as a comparatively homogeneous reference system for evaluating methodological influences, as it does not exhibit the local variability introduced by glass fibers at the mesoscale. In contrast, the indentation response of PBT~GF30 reflects the apparent local behavior of a heterogeneous material volume, incorporating varying contributions from the polymer matrix, glass fibers, and adjacent fiber-matrix regions~\cite{Rauter_habil2023}.

Additionally, Figure~\ref{fig:injection_molding_flow} illustrates the process-induced microstructural heterogeneity of PBT~GF30 by presenting the geometry of the injection-molded tensile specimen, schematic representations of melt-flow conditions, a representative optical micrograph of the reinforced material (PBT~GF30) as well as the coordinate system definition used throughout this work, where the longitudinal specimen direction corresponds to the $y$-axis, the transverse in-plane direction with the $x$-axis, and the specimen-thickness direction with the $z$-axis. Therefore, during injection molding, the melt primarily flows along $y$-axis, which is the longitudinal specimen direction. The longitudinal flow-velocity gradient across the thickness direction $z$ produces a nonuniform shear-rate distribution. Combined with the fountain-flow mechanism at the advancing melt front, these flow conditions result in a layered microstructure with spatially varying fiber orientations~\cite{Eyerer2008, Huang2020}.


\subsection{Specimen extraction and surface preparation}\label{subsec:specimen_preparation}

\begin{figure}[t]
   \centering
   \def\svgwidth{0.99\columnwidth}
    \graphicspath{{images/}}
    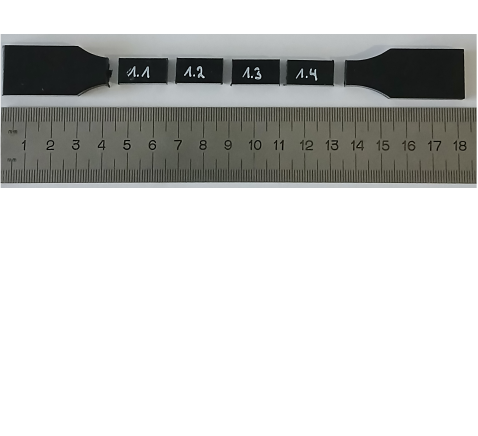
    \caption{Extraction and preparation of specimens for instrumented indentation. The upper schematic illustrates the extraction of specimen sections from the gage region of an injection-molded tensile specimen and the orientation of the coordinate system relative to the melt-flow direction. The lower photographs show the extraction positions, the separated specimen sections, and the embedded samples prepared for subsequent indentation measurements.}
    \label{fig:specimen_extraction}
\end{figure}

For the instrumented indentation testing, small sections were extracted from the uniform gage region of the injection-molded tensile specimens, as illustrated in Figure~\ref{fig:specimen_extraction}. The extracted sections were identified according to their longitudinal position to retain information about their location within the original specimen.

The extracted sections were oriented such that the $x$--$z$ cross-section, perpendicular to the principal melt-flow direction, formed the surface investigated by instrumented indentation. Consequently, the indentation load was applied parallel to the longitudinal $y$-direction of the original tensile specimen.

The extracted sections were subsequently embedded and prepared using a sequential metallographic grinding and polishing procedure. Coarse material removal was followed by fine silicon carbide (SiC) grinding up to P2500~\cite{ELSafty2012}. The surfaces were then polished with progressively finer media, including \SI{9}{\micro\metre} and \SI{3}{\micro\metre} diamond suspensions, and finished with a final submicrometer polishing step~\cite{Oliver1992, Shirasu2020}. After each preparation stage, abrasive residues and material debris were removed to prevent contamination of subsequent polishing steps.

This preparation procedure was chosen to produce a planar and sufficiently smooth cross-section suitable for both contact detection and optical assessment of the resulting indentation imprints. Preliminary measurements on surfaces finished after P2500 grinding were retained for the methodological investigation described in Section~\ref{sec:methodological_assessment}. Final spatial measurements on PBT~GF0 and PBT~GF30 were conducted exclusively on fully polished cross-sections.

\subsection{Instrumented indentation system}\label{subsec:indentation_setup}

\begin{figure}[h]
   \centering
   \def\svgwidth{0.99\columnwidth}
    \graphicspath{{images/}}
    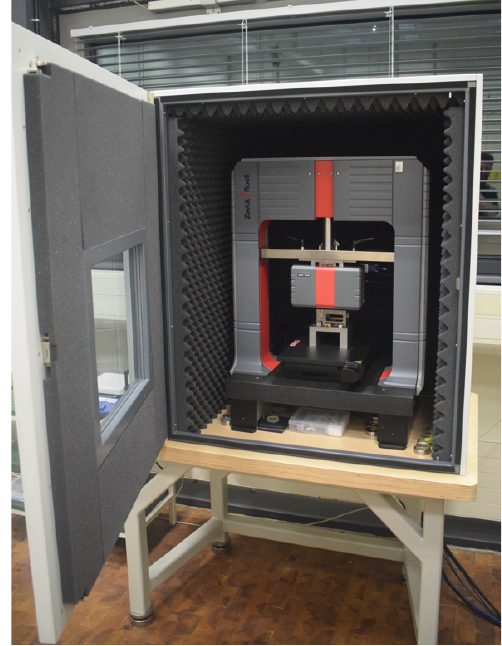
    \caption{The ZHN instrumented indentation system (ZwickRoell/ASMEC) is installed within a thermally and acoustically insulated enclosure for experimental investigations. The enclosure is depicted with its access door open, revealing the indentation instrument and measuring assembly. This enclosure minimizes environmental disturbances during testing.}
    \label{fig:Setup_ZHN_indentation}
\end{figure}

All indentation experiments were conducted using an instrumented ZHN indentation system supplied by ZwickRoell in collaboration with ASMEC. Figure~\ref {fig:Setup_ZHN_indentation} presents the experimental setup, which includes the instrument and its protective enclosure. Table~\ref {tab:head} provides the specifications of the \SI{2}{\newton} normal-force unit measuring head.

For the indentation testing the system was equipped with a Berkovich diamond tip. The area function of the indenter and the compliance of the measurement system were calibrated before the experimental investigations. The specimens were mounted with the prepared cross-section perpendicular to the indentation axis. Additionally, thermal drift was addressed by combining environmental shielding with thermal stabilization prior to testing. The ZHN indentation system was operated within its thermally and acoustically insulated enclosure, which reduces exposure of the measurement assembly to ambient temperature fluctuations, air currents, and acoustic disturbances. By attenuating changes in the local thermal environment, the enclosure limits temperature-induced dimensional changes within the displacement measurement chain that could otherwise contribute to the recorded indentation depth. Both the specimens and the indentation system were allowed to reach thermal equilibrium in a temperature-controlled laboratory environment for an extended period prior to testing \cite{Jakes2021, Meng2013}. This thorough thermal stabilization ensured that the thermal drift rate was negligible. Therefore, explicit thermal drift correction was omitted, which is a methodologically established simplification when drift-induced displacement is insignificant compared to the large maximum indentation depths of up to several micrometers examined in this study \cite{Ngan2002}. However, to eliminate the effects of surface roughness and instrument compliance, a zero-point correction was automatically applied to the raw load-displacement data. The exact point of initial surface contact was determined by fitting a power-law function, that is based on the Hertzian contact model, to the initial segment of the loading curve. 

\begin{table}[t]
\centering
\small
\caption{Specification of the used \SI{2}{\newton} normal force unit measuring head provided by ZwickRoell~\cite{ZwickRoell} and the official calibration protocol.}
\label{tab:head}
\begin{tabular}{lll}
\toprule
Quantity & Value & Unit\\
\midrule
Maximal test load $F_{\max}$& $\pm \ 2$ & N\\
Minimal test load $F_{\min}$& approx. $0.2$ & mN\\
Digital resolution force & $\le \ 0.02$ & \si{\micro\newton}\\[1ex]
Calibration information & \multicolumn{2}{l}{Expanded measurement}\\
& \multicolumn{2}{l}{uncertainty of approx. $\pm 0.30 \%$}\\
\midrule
Maximal displacement & $\pm \ 200$ & \si{\micro\meter}\\
Digital resolution displacement & $\le \ 0.3$ & nm\\[1ex]
Calibration information &  \multicolumn{2}{l}{Resolution up to \SI{20}{\micro\meter} of \SI{0.9}{\nano\meter}}\\
& \multicolumn{2}{l}{and up to \SI{200}{\micro\meter} of \SI{1.1}{\nano\meter};}\\
&   relative measure-\\
& ment uncertainty of $\pm 0.30 \%$\\
\bottomrule
\end{tabular}
\end{table}

\subsection{Confocal surface and indentation-imprint analysis}
\label{subsec:confocal_analysis}

Supplementary confocal measurements were performed using a KEYENCE VK-X200 series confocal laser scanning microscope to characterize the effects of surface preparation and the residual geometry of selected indentation imprints. Representative indentation sites on P2500-ground and polished PBT~GF0 surfaces were examined after unloading to identify local surface elevations associated with pile-up and to evaluate the surface topography surrounding the residual indentation cavity.

Confocal measurements were performed at a magnification of $150\,\times$, which provided sufficient spatial resolution to resolve the triangular Berkovich imprint and the surrounding surface topography. Surface profiles across the indentation cavity were analyzed to detect local elevations or depressions relative to the undeformed surface. Surface roughness was quantified using the arithmetic mean roughness $R_\mathrm{a}$ and the mean roughness depth $R_\mathrm{z}$, both calculated from multiple line profiles outside the residual indentation region to ensure that contact detection was not influenced by residual topography \cite{Gross2013Nanoindentation, Hardiman2016}.

The confocal measurements provided supplementary geometrical assessment and were not used to adjust the projected contact area in the Oliver–Pharr evaluation. Consequently, the profile measurements support the interpretation of the indentation results but do not constitute an independent contact-area correction method. This approach is particularly robust for polymer systems, since rapid and pronounced viscoelastic recovery immediately after indenter unloading significantly alters the residual profile depth, rendering direct area corrections from post-indentation topography measurements highly problematic \cite{Hardiman2016}.

\section{Methodological assessment using unreinforced PBT~GF0}
\label{sec:methodological_assessment}

The initial methodological investigations were conducted on PBT~GF0 to assess the influence of experimental and evaluation parameters while limiting the additional variability introduced by the reinforcing phase. First, the effects of indentation spacing and surface preparation were examined, supported by confocal measurements of surface roughness and residual indentation geometry. Subsequently, controlled comparisons were used to assess the influence of maximum load, holding duration, and unloading duration. The sensitivity of the apparent indentation modulus to the unloading fitting interval was evaluated separately. Finally, the influence of loading time on the apparent modulus and the holding-period depth increment was investigated. These investigations informed the selection of a common protocol for the subsequent spatial comparison of PBT~GF0 and PBT~GF30. Therefore, each investigation addresses a specific aspect of the experimental or analytical procedure. Table~\ref{tab:methodological_investigations} summarizes these objectives and their role in the subsequent spatial comparison of PBT~GF0 and PBT~GF30.

\begin{table*}[t]
    \centering
    \caption{Objectives and methodological roles of the conducted preliminary investigations to assess selected influences on the indentation response and support the subsequent comparison of spatially distributed measurements.}
    \label{tab:methodological_investigations}
    \small
    \begin{tabular*}{\textwidth}{    
    @{\extracolsep{\fill}}
    *{4}{>{\raggedright\arraybackslash}
         p{\dimexpr(\textwidth-8\tabcolsep)/4\relax}}
    @{}}
        \toprule
        \textbf{Methodological aspect}
        & \textbf{Objective}
        & \textbf{Experimental or analytical approach}
        & \textbf{Role in the subsequent analysis} \\
        \midrule
        Interaction between neighboring indentations (Section~\ref{subsec:spacing})
        & Assess whether the measured response depends on the distance from a preceding indentation
        & Three PBT~GF0 sequences with increasing spacing at \SI{1750}{\milli\newton}
        & Support the selection of the point-to-point spacing \\ \addlinespace

        Surface condition (Section~\ref{subsec:surface_preparation})
        & Assess the influence of preparation on the measured response and residual surface geometry  
        & Comparison of P2500-ground and polished PBT~GF0 surfaces, supplemented by confocal measurements
        & Establish a consistent surface condition and support interpretation of the residual imprints \\ \addlinespace

        Maximum indentation load (Section~\ref{subsec:parameter_sensitivity_load})
        & Assess the sensitivity of indentation depth and apparent modulus to maximum load
        & Controlled comparison of \SI{500}{} and \SI{1000}{\milli\newton} measurements
        & Distinguish changes in penetration depth from changes in evaluated modulus \\ \addlinespace

        Holding and unloading durations (Section~\ref{subsec:parameter_sensitivity})
        & Assess the influence of temporal loading parameters on the measured response
        & Controlled comparisons of holding and unloading durations at \SI{500}{\milli\newton}
        & Support the selection and interpretation of the temporal test conditions \\ \addlinespace

        Unloading evaluation (Section~\ref{subsec:unloading_evaluation})
        & Assess the dependence of the apparent modulus on the fitting interval
        & Evaluation of the same curves using broad and initial unloading intervals under three holding/unloading conditions
        & Identify sensitivity associated with the unloading-curve evaluation \\ \addlinespace

        Loading duration at final load level of \SI{1750}{\milli\newton} (Section~\ref{subsec:loading_time})
        & Assess the influence of the preceding ramp on the subsequent holding response
        & Comparison of four loading durations with fixed maximum load, holding time, and unloading time
        & Assess the loading-history dependence of $E_{\mathrm{ind}}$ and $\Delta h_{\mathrm{cr}}$ \\ \addlinespace

        \bottomrule
    \end{tabular*}
\end{table*}

\subsection{Minimum indentation spacing}\label{subsec:spacing}
\subsubsection{Experimental setup and measurement procedure}

Spatially resolved measurements are required to characterize the variability induced by the microstructural heterogeneity. In order to ensure the independence of these measurements across the surface the minimum distance between consecutive indentation positions was investigated first. Since the minimum distance is directly linked to the indentation area and hence to the indentation depth this analysis was conducted at a maximum load of $P_\mathrm{max}=\SI{1750}{\milli\newton}$, which equals $87.5 \%$ of the maximal operational load of \SI{2000}{\milli\newton}. 

Three measurement sequences were performed on PBT~GF0, with the point-to-point distance systematically varied from $\SI{50}{\micro\metre}$ to $\SI{600}{\micro\metre}$ in $\SI{50}{\micro\metre}$ increments, as depicted in Figure~\ref{fig:spacing_sequences}. The investigation employed a loading time of $\SI{10}{\second}$, a holding time of $\SI{10}{\second}$, and an unloading time of $\SI{4}{\second}$ as well as a polished measurement surface.

\subsubsection{Minimum indentation spacing measurement results}

\begin{figure}[!h]
   \centering
   \def\svgwidth{0.99\columnwidth}
    \graphicspath{{images/}}
    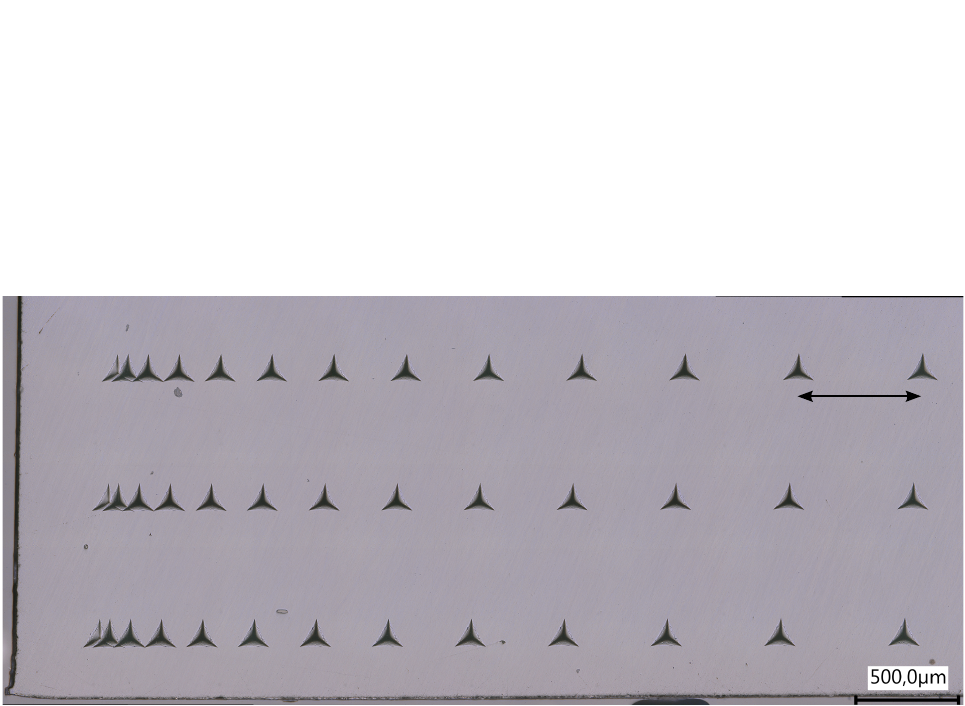
    
    \vspace{1.5em}

    \includegraphics[width=1\columnwidth]{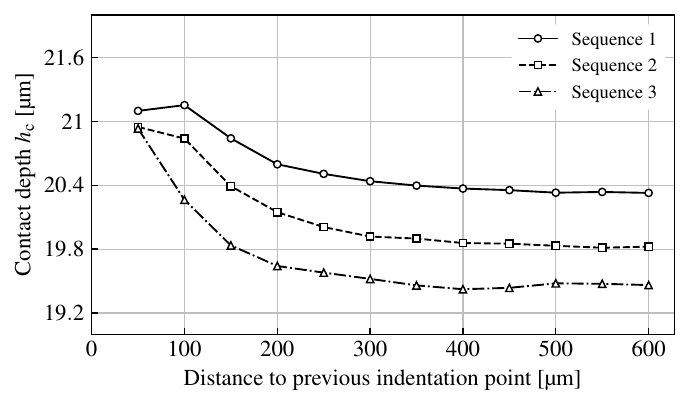}
    \caption{Experimental investigation of indentation spacing in PBT~GF0 at $P_\mathrm{max}=\SI{1750}{\milli\newton}$. Top: Optical image of the $x$--$z$ specimen cross-section showing the locations of the indentation sequences. Middle: Magnified view of three indentation sequences with progressively increasing distances between consecutive indentations. Bottom: Contact depth $h_\mathrm{c}$ plotted as a function of the distance to the preceding indentation for the three measurement sequences. The loading, holding, and unloading times were $\SI{10}{\second}$, $\SI{10}{\second}$, and $\SI{4}{\second}$, respectively.}
    \label{fig:spacing_sequences}
\end{figure}

Figure~\ref{fig:spacing_sequences} also presents the corresponding contact-depth response. The optical images display the specimen cross-section and three indentation sequences with systematically increasing distances between adjacent positions. The accompanying diagram shows the contact depth $h_\mathrm{c}$ as a function of the distance to the preceding indentation. At smaller distances, contact depth demonstrated a marked dependence on indentation spacing, indicating an effect from the deformation field of the previous indentation. While the absolute contact depths varied among the three sequences, their sensitivity to indentation spacing diminished as the distance increased.

With increasing indentation spacing, the contact depth approached a nearly constant value within each measurement sequence. Beyond approximately $\SI{400}{\micro\metre}$, only minor variations were observed, indicating that the influence of the preceding indentation became negligible compared to the remaining variation within individual sequences. Therefore, a spacing of approximately $\SI{400}{\micro\metre}$ was established as a practical lower limit under the experimental conditions investigated.

\subsubsection{Discussion and spacing definition for subsequent measurements}

The spacing investigation demonstrated that, under the experimental conditions, contact depth was largely independent of the distance to the preceding indentation for separations greater than $\SI{400}{\micro\metre}$. Accordingly, a point-to-point distance of $\SI{500}{\micro\metre}$ was chosen for the final spatial indentation measurements to ensure an additional margin beyond the identified lower limit.

\begin{figure}[t]
   \centering
   \def\svgwidth{0.99\columnwidth}
    \graphicspath{{images/}}
    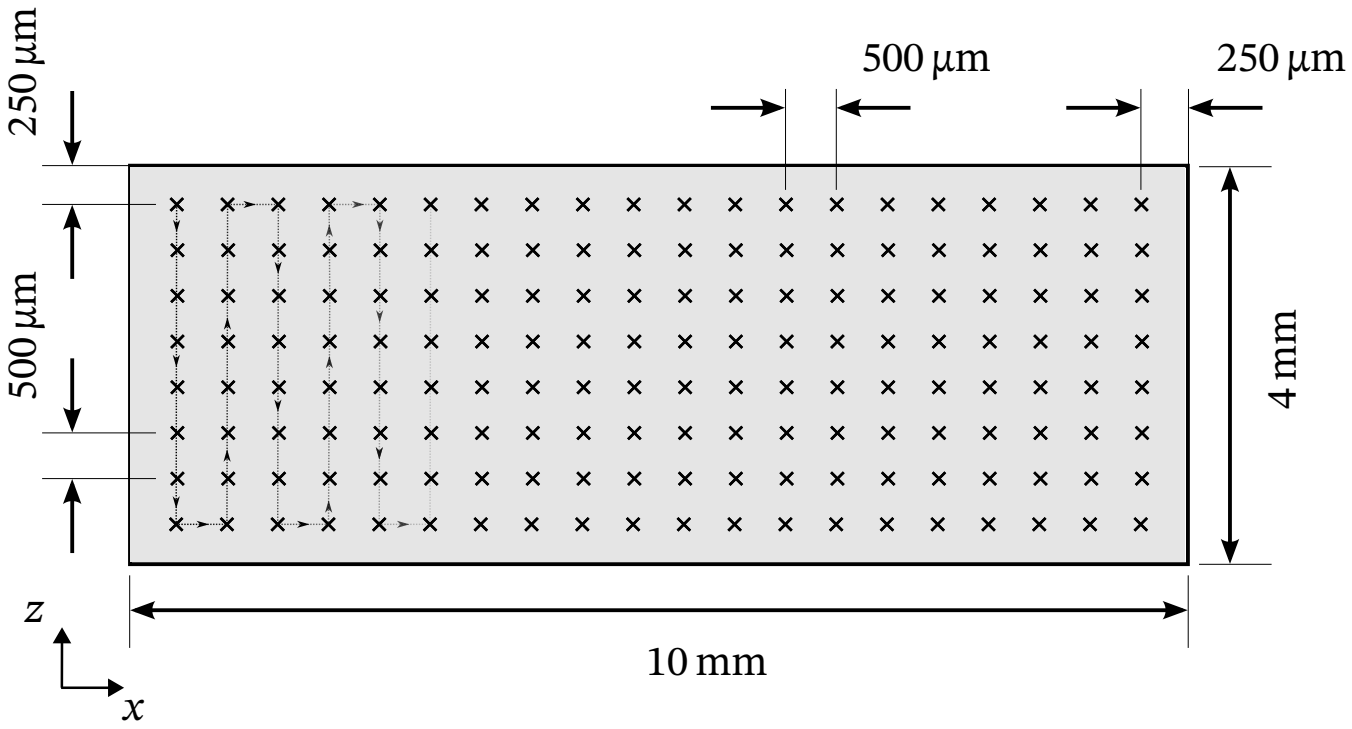
    \caption{Schematic arrangement of the spatial indentation grid on the polished $x$--$z$ cross-section of the injection-molded tensile specimen. A total of $160$ indentation positions were arranged in $8$ rows and $20$ columns, with a nominal point-to-point distance of $\SI{500}{\micro\metre}$ and a boundary offset of $\SI{250}{\micro\metre}$. The nominal cross-sectional dimensions were $\SI{10}{\milli\metre}\times\SI{4}{\milli\metre}$.}
    \label{fig:indentation_grid}
\end{figure}

Figure~\ref{fig:indentation_grid} presents a schematic of the measurement grid. The grid comprised a regular arrangement of $8\times20$ indentation positions distributed across the polished $x$--$z$ specimen cross-section, yielding $160$ measurement points. The nominal cross-sectional dimensions were $\SI{10}{\milli\metre}$ in the $x$-direction and $\SI{4}{\milli\metre}$ in the $z$-direction. A uniform point-to-point distance of $\SI{500}{\micro\metre}$ and a boundary offset of $\SI{250}{\micro\metre}$ were implemented to ensure spatial coverage and to avoid positions adjacent to the specimen edges.

The chosen spacing also surpasses commonly reported empirical recommendations, which are typically expressed as multiples of the maximum indentation depth~\cite{Hintsala2018, Jiang2020, SudharshanPhani2019, Rossi2023}. For the final PBT~GF0 measurements, the mean maximum indentation depth was approximately $\SI{24.16}{\micro\metre}$, resulting in a spacing-to-depth ratio of about $21$. However, geometric spacing criteria alone do not eliminate the possibility of mechanical interactions, especially in viscoelastic and heterogeneous polymer systems.

\subsection{Influence of surface preparation and residual indentation geometry}\label{subsec:surface_preparation}
\subsubsection{Surface preparation details and measurement procedure}

Maintaining surface roughness below critical thresholds is crucial to avoid initial contact-detection errors and to ensure that local spatial scatter accurately represents material variability rather than artifacts introduced by surface topography \cite{Hardiman2016, McAllister2012}. Therefore, in a second step the influence of surface preparation on indentation measurements and residual imprints was investigated by comparing two PBT~GF0 specimens, one finished by P2500 grinding and the other subjected to additional polishing. The measurements follow the defined grid of $\SI{500}{\micro\meter} \times \SI{500}{\micro\meter}$ with an edge distance of $\SI{250}{\micro\meter}$, resulting in a total of 160 measurement points, as shown in Figure~\ref{fig:indentation_grid}. Additionally, the measurement surfaces were analyzed by confocal measurements after the indentation measurements including the surface roughness and the residual indentation geometry.

\begin{figure}[!h]
   \centering
   \def\svgwidth{0.99\columnwidth}
    \graphicspath{{images/}}
    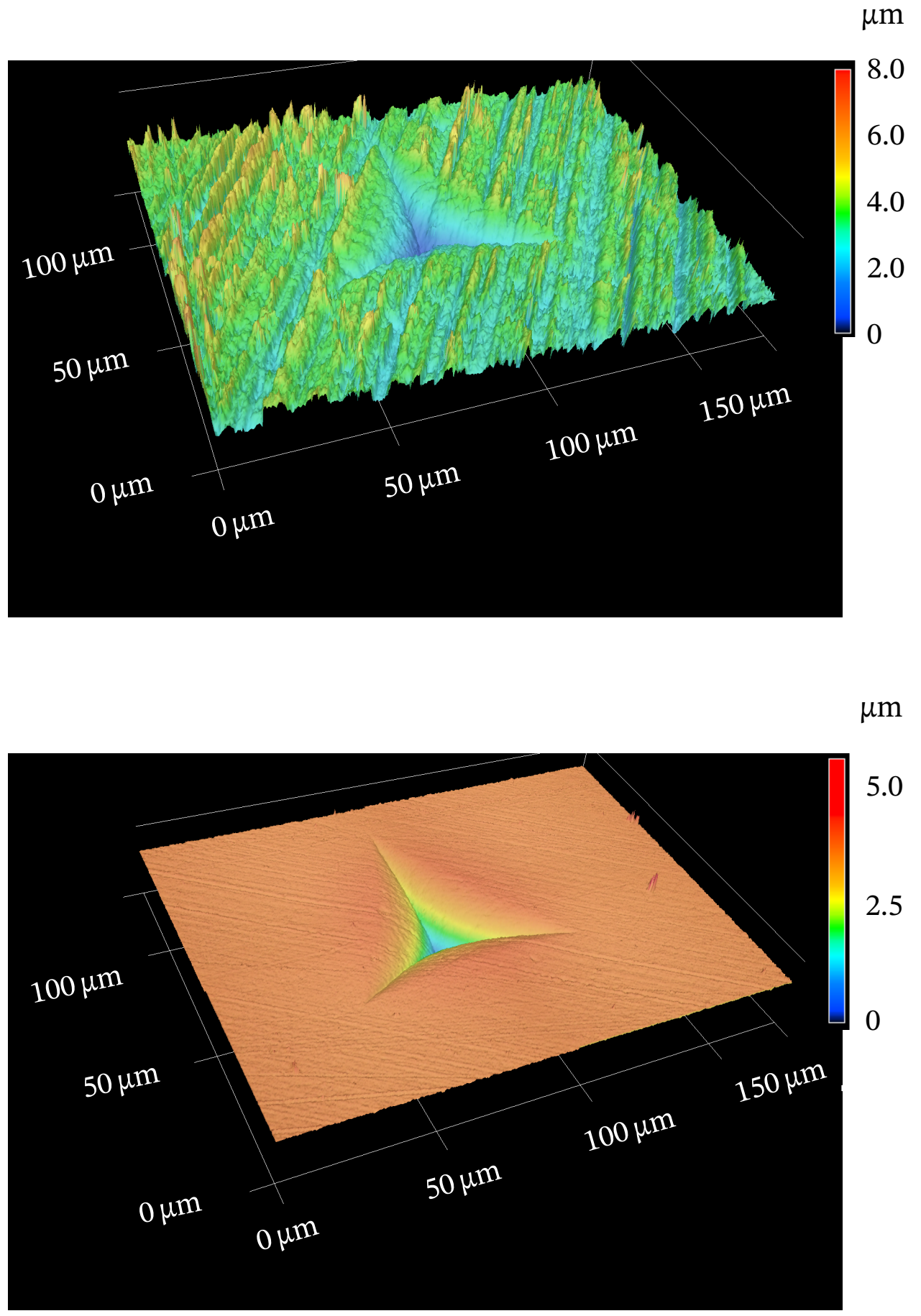
    \caption{Comparison of the residual indentation geometry on PBT~GF0 surfaces with different preparation conditions, obtained using a KEYENCE VK-X200 series confocal laser scanning microscope. Top: Three-dimensional surface topography of a P2500-ground specimen, showing pronounced surface irregularities that complicate the identification of the residual indentation imprint. 
    Bottom: Corresponding three-dimensional surface topography of a polished specimen, allowing clearer identification of the residual Berkovich imprint and the surrounding surface deformation.}
    \label{fig:confocal_surface_preparation}
\end{figure}

Because initial contact detection occurs at relatively low loads, surface topography can influence the determination of the displacement origin. To evaluate the influence of surface preparation, both polished and P2500-ground specimens were tested under identical loading conditions, with $P_\mathrm{max}=\SI{500}{\milli\newton}$, $t_\mathrm{load}=\SI{6}{\second}$, $t_\mathrm{hold}=\SI{20}{\second}$, and $t_\mathrm{unload}=\SI{2}{\second}$. The polished and P2500-ground conditions correspond to configurations No.~1 and No.~2, respectively. Configuration No.~1 was used as the reference for comparisons involving surface preparation, maximum load, and unloading duration, while configuration No.~2 served as the reference for the holding-time comparison on P2500-ground surfaces. All apparent indentation moduli were calculated using the unloading interval from $98$ to $40\,\%$ of $P_\mathrm{max}$. The relevant measurement configurations and results are presented in Table~\ref{tab:selected_parameter_comparisons}. Figure~\ref{fig:confocal_surface_preparation} shows representative residual indentation imprints for the two surface conditions.

\begin{table*}[!t]
\centering
\caption{Comparisons of selected experimental parameters for PBT~GF0 are presented. Configuration No.~1 is used as the reference for surface preparation, maximum load, and unloading duration, while configuration No.~2 is the reference for holding-time comparisons on P2500-ground surfaces. All apparent indentation moduli are evaluated using the $98$--$40\,\%$ unloading interval of $P_\mathrm{max}$. The loading-time investigation is conducted separately at $P_\mathrm{max}=\SI{1750}{\milli\newton}$ with a holding duration of $\SI{300}{\second}$. Values are presented as the mean $\pm$ one standard deviation.}
\label{tab:selected_parameter_comparisons}
\begin{tabular*}{\textwidth}
{@{\extracolsep\fill}llcccccccc@{\extracolsep\fill}}
\toprule
\textbf{Comparison} &
\textbf{Configuration} &
$\boldsymbol{P_\mathrm{max}}$ &
$\boldsymbol{t_\mathrm{load}}$ &
$\boldsymbol{t_\mathrm{hold}}$ &
$\boldsymbol{t_\mathrm{unload}}$ &
\textbf{Surface} &
\textbf{Measurement} &
$\boldsymbol{E_\mathrm{ind}}$ &
$h_\mathrm{c}$ \\
&
&
$[\mathrm{mN}]$ &
$[\mathrm{s}]$ &
$[\mathrm{s}]$ &
$[\mathrm{s}]$ &
\textbf{preparation}& \textbf{points}&
$[\SI{}{\giga\pascal}]$ &
$[\SI{}{\micro\metre}]$ \\
\midrule

Reference
& No.~1
& 500 & 6 & 20 & 2 & polished & 160
& $3.00\pm0.18$
& $10.95\pm0.25$ \\

\addlinespace

Surface preparation
& No.~2
& 500 & 6 & 20 & 2 & P2500-ground & 160
& $2.96\pm0.21$
& $11.25\pm0.44$ \\

\addlinespace

Load
& No.~3
& 1000 & 6 & 20 & 2 & polished & 160
& $2.88\pm0.22$
& $15.58\pm0.44$ \\

\addlinespace

Holding time
& No.~4
& 500 & 6 & 1 & 2 & P2500-ground & 160
& $2.84\pm0.23$
& $10.74\pm0.81$ \\

& No.~5
& 500 & 6 & 10 & 2 & P2500-ground & 160
& $2.95\pm0.22$
& $11.17\pm0.79$ \\

\addlinespace

Unloading time
& No.~6
& 500 & 6 & 1 & 20 & P2500-ground & 160
& $3.01\pm0.20$
& $10.85\pm0.56$ \\

& No.~7
& 500 & 6 & 20 & 20 & polished & 160
& $2.75\pm0.19$
& $11.09\pm0.28$ \\

\addlinespace
\midrule
\addlinespace
\multicolumn{10}{c}{High-load reference$^{\mathrm{a}}$}\\
\addlinespace
\midrule

Loading time
& No.~8
& 1750 & 1 & 300 & 4 & polished & 40
& $2.89\pm0.12$
& $21.31\pm0.37$ \\

& No.~9
& 1750 & 3 & 300 & 4 & polished & 40
& $3.02\pm0.12$
& $20.56\pm0.57$ \\

& No.~10$^{\mathrm{b}}$
& 1750 & 6 & 300 & 4 & polished & 40
& $2.91\pm0.09$
& $21.25\pm0.31$ \\

& No.~11
& 1750 & 10 & 300 & 4 & polished & 40
& $3.03\pm0.07$
& $20.51\pm0.19$ \\

\bottomrule
\end{tabular*}

\begin{tablenotes}
\item[$^{\mathrm{a}}$] The $1750~\mathrm{mN}$ measurements were obtained using the independently established final long-hold protocol with a holding time of $\SI{300}{\second}$, and an unloading time of $\SI{4}{\second}$. This condition is therefore included as an additional high-load consistency reference and not as a third level of the controlled load comparison.
\item[$^{\mathrm{b}}$] Final configuration used for the spatially resolved measurements.
\end{tablenotes}
\end{table*}

\subsubsection{Indentation and confocal measurement results}

Although the mechanical response exhibited only minor differences between these methods, substantial variations were observed in surface geometry and the clarity of residual indentation imprints.

Initial confocal measurements showed that the mean roughness-depth values $R_\mathrm{z}$ ranged from approximately $\SI{1.5}{\micro\metre}$ to $\SI{3.0}{\micro\metre}$ for the P2500-ground specimen. After polishing, $R_\mathrm{z}$ decreased to approximately $\SI{0.2}{\micro\metre}$ to $\SI{0.4}{\micro\metre}$. The pronounced topographical variations on ground surfaces hindered the distinction between the residual indentation cavity, surrounding pile-up, and undeformed specimen surface. In contrast, polished surfaces allowed for clearer identification of the residual Berkovich imprint and adjacent surface elevations~\cite{Huszar2024}.

\begin{figure}[t]
   \centering
   \def\svgwidth{0.99\columnwidth}
    \graphicspath{{images/}}
    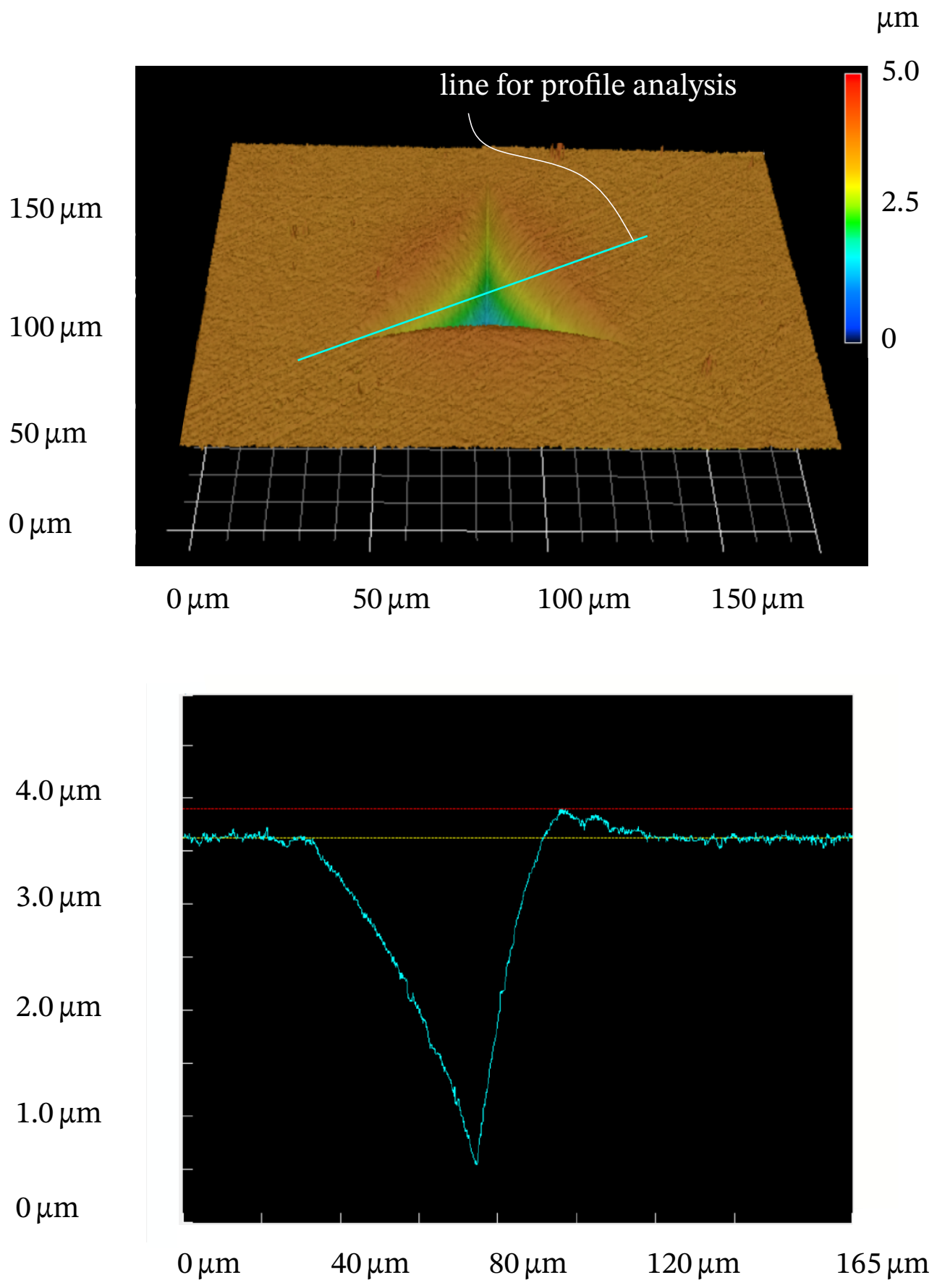
    \caption{Confocal characterization of the residual indentation geometry of polished PBT~GF0 following indentation at $P_\mathrm{max}=\SI{500}{\milli\newton}$, in accordance with configuration No.~1 in Table~\ref{tab:selected_parameter_comparisons}. Top: Three-dimensional surface topography of a representative Berkovich indentation imprint obtained using a KEYENCE VK-X200 series confocal laser scanning microscope. The indicated line defines the position of the profile evaluation. Bottom: Corresponding surface height profile illustrating the residual indentation cavity and adjacent elevated material regions associated with pile-up.}
    \label{fig:confocal_pileup_profile}
\end{figure}

The residual indentation geometry of polished PBT~GF0 was examined in greater detail using confocal laser scanning microscopy. Figure~\ref{fig:confocal_pileup_profile} displays a representative three-dimensional surface topography of a residual Berkovich indentation imprint produced at $P_\mathrm{max}=\SI{500}{\milli\newton}$, along with the corresponding surface height profile. The results indicate a distinct residual indentation cavity and elevated material adjacent to the imprint, which provides direct experimental evidence of pile-up following unloading.

For the representative polished PBT~GF0 imprints examined, residual indentation depths ranged from approximately $\SI{2.6}{\micro\metre}$ to $\SI{4.5}{\micro\metre}$, while local pile-up heights were approximately $\SI{0.2}{\micro\metre}$ to $\SI{0.4}{\micro\metre}$. These findings indicate that elastic sink-in alone does not fully account for the residual deformation surrounding the indentation cavity. As a result, the actual projected contact area during indentation may differ from that estimated using the conventional Oliver--Pharr contact-depth relation~\cite{Oliver1992, Cheng2004, Hardiman2012, SarkarPantNanavati2024}. Confocal measurements characterize the residual surface geometry after unloading, which may be influenced by subsequent viscoelastic recovery. Therefore, the measured pile-up heights were used for qualitative assessment rather than quantitative correction of the projected contact area.

The effect of surface preparation on the mean apparent indentation modulus was relatively minor. Under otherwise identical experimental conditions, the controlled comparison yielded $E_\mathrm{ind}=\SI{2.96}{\giga\pascal}$ for the P2500-ground surface (configuration No.~$2$) and $E_\mathrm{ind}=\SI{3.00}{\giga\pascal}$ (configuration No.~$1$) for the polished surface, a difference of $\SI{0.04}{\giga\pascal}$. The standard deviation decreased from $\SI{0.21}{\giga\pascal}$ to $\SI{0.18}{\giga\pascal}$ (approx. $\SI{14}{\percent}$). These results indicate that polishing reduced the variability in the measured indentation modulus while causing only a minor change in its mean value. Similarly, the mean indentation depth $h_\mathrm{c}$ changed only from $11.25\pm \SI{0.44}{\micro\metre}$ to $10.95\pm\SI{0.25}{\micro\metre}$. In this case, polishing reduced the standard deviation by approximately $\SI{43}{\percent}$. Thus, the primary benefit of polishing was a reduction in measurement scatter rather than a systematic shift in the mean response~\cite{Huszar2024}. This finding is consistent with the reduced surface roughness and improved identification of residual imprints. All measurement results are summarized in Table~\ref{tab:selected_parameter_comparisons}.

In summary, surface polishing created more favorable conditions for spatially resolved indentation measurements by reducing surface roughness and improving the identification of residual indentation geometry. Because the observed pile-up was not included in a pointwise correction of the projected contact area, the resulting indentation modulus values are consistently interpreted as apparent indentation moduli. This limitation influences the interpretation of their absolute magnitude and should be considered when comparing unreinforced and reinforced material systems.

\subsection{Sensitivity to indentation load}
\label{subsec:parameter_sensitivity_load}

A complementary set of PBT~GF0 measurements was conducted to evaluate the sensitivity of the indentation response to maximum load utilizing the already presented measurement setup with a total of $160$ measurement points where the maximum load $P_\mathrm{max}$ was increased from $\SI{500}{\milli\newton}$ (configuration No.~$1$) to $\SI{1000}{\milli\newton}$ (configuration No.~$3$). 

The most significant response to an increase in the maximum indentation load was observed in the indentation depth. Under otherwise identical conditions, increasing $P_\mathrm{max}$ from $\SI{500}{\milli\newton}$ to $\SI{1000}{\milli\newton}$ raised the mean indentation depth from $\SI{10.95}{\micro\metre}$ to $\SI{15.58}{\micro\metre}$ (approximately $\SI{42}{\percent}$), as expected. In contrast, the apparent indentation modulus changed only from $\SI{3.00}{\giga\pascal}$ to $\SI{2.88}{\giga\pascal}$, a difference of approximately $\SI{4}{\percent}$. Thus, in this controlled comparison, the increase in maximum load mainly affected the magnitude of the indentation rather than the local stiffness evaluated.

The associated standard deviations corroborate this interpretation. For $E_\mathrm{ind}$, the standard deviation increased moderately from $\SI{0.18}{\giga\pascal}$ to $\SI{0.22}{\giga\pascal}$. For indentation depth, the increase from $\SI{0.25}{\micro\metre}$ to $\SI{0.44}{\micro\metre}$ reflects the larger absolute displacement at higher indentation load. Therefore, increased penetration was accompanied by greater absolute depth scatter, while the mean apparent indentation modulus remained nearly unchanged.

This outcome confirms that the apparent modulus is largely independent of peak force within this load range. This load independence is physically expected for geometrically self-similar indenters (such as the Berkovich tip) once the indentation depth is sufficiently large to overcome tip blunting and near-surface artifacts, typically occurring at contact depths exceeding $\SI{100}{\nano\meter}$ \cite{Jakes2018, Nix1998}. However, these results do not demonstrate complete load independence.

\subsection{Sensitivity to holding and unloading duration}
\label{subsec:parameter_sensitivity}

The effects of holding and unloading durations were examined using PBT~GF0 measurements at $P_\mathrm{max}=\SI{500}{\milli\newton}$ with a constant loading duration of $\SI{6}{\second}$. For the holding-time comparison, configurations No.~4 and No.~5, with holding durations of $\SI{1}{\second}$ and $\SI{10}{\second}$, respectively, were compared to configuration No.~2, which used a holding duration of $\SI{20}{\second}$. All three configurations were performed on P2500-ground surfaces with an unloading duration of $\SI{2}{\second}$. The influence of unloading duration was assessed through two complementary comparisons: configurations No.~4 and No.~6 on P2500-ground surfaces with a holding duration of $\SI{1}{\second}$, and configurations No.~1 and No.~7 on polished surfaces with a holding duration of $\SI{20}{\second}$. The results are presented in Table~\ref{tab:selected_parameter_comparisons}.

An increase in holding duration from $\SI{1}{\second}$ to $\SI{20}{\second}$ resulted in a rise in the mean apparent indentation modulus from $\SI{2.84}{\giga\pascal}$ to $\SI{2.96}{\giga\pascal}$, representing approximately $4.2\,\%$. At the intermediate holding duration of $\SI{10}{\second}$, the mean apparent indentation modulus was $\SI{2.95}{\giga\pascal}$. The corresponding mean contact depths $h_\mathrm{c}$ were $\SI{10.74}{\micro\metre}$, $\SI{11.17}{\micro\metre}$, and $\SI{11.25}{\micro\metre}$ for holding durations of $\SI{1}{\second}$, $\SI{10}{\second}$, and $\SI{20}{\second}$, respectively. These findings demonstrate a moderate variation in the indentation response with holding duration. However, as the compared configurations originated from different preliminary measurement series, the observed differences should be interpreted descriptively rather than attributed solely to the holding-time effect.

For P2500-ground PBT~GF0, extending the unloading duration from $\SI{2}{\second}$ (No.~4) to $\SI{20}{\second}$ (No.~6) at a fixed holding duration of $\SI{1}{\second}$ increased the mean apparent indentation modulus from $\SI{2.84}{\giga\pascal}$ to $\SI{3.01}{\giga\pascal}$, an increase of approximately $6.0\,\%$. The mean contact depth also increased slightly from $\SI{10.74}{\micro\metre}$ to $\SI{10.85}{\micro\metre}$. Conversely, on polished surfaces at a holding duration of $\SI{20}{\second}$ (No.~1 and No.~7), the mean apparent indentation modulus decreased from $\SI{3.00}{\giga\pascal}$ to $\SI{2.75}{\giga\pascal}$, a reduction of approximately $8.3\,\%$. The associated mean contact depth increased from $\SI{10.95}{\micro\metre}$ to $\SI{11.09}{\micro\metre}$.

These contrasting trends indicate that the apparent indentation modulus displays varying sensitivities to unloading duration under the tested conditions. Notably, differences in preceding holding duration and surface preparation between the two comparisons preclude unambiguous attribution of the observed trends to a single experimental parameter. Continued viscoelastic deformation and recovery during unloading may also affect the unloading curvature and the resulting contact stiffness~\cite{Cheng2004, Hardiman2016, Hochstetter1999}. These results underscore the necessity of maintaining consistent temporal testing and evaluation conditions when comparing indentation responses.

\subsection{Influence of the unloading evaluation interval}
\label{subsec:unloading_evaluation}

The sensitivity of the apparent indentation modulus to the unloading fitting interval was assessed to quantify the effect of ongoing viscoelastic deformation immediately after peak load~\cite{Hardiman2016, Hochstetter1999}. This evaluation involved analyzing identical force--displacement responses over the $98$--$40\,\%$ and $100$--$90\,\%$ intervals of $P_\mathrm{max}$. Configurations No.~4, No.~5 and No.~6 were included in the analysis, all performed on P2500-ground PBT~GF0 surfaces at $P_\mathrm{max}=\SI{500}{\milli\newton}$. The corresponding mean values and standard deviations are provided in Table~\ref{tab:unloading_interval}.

\begin{table}[!ht]
\centering
\small
\caption{Influence of the unloading fitting interval on the apparent indentation modulus of PBT~GF0 at $P_\mathrm{max}=\SI{500}{\milli\newton}$. All measurements were performed on P2500-ground surfaces. Values are reported as mean $\pm$ one standard deviation.}
\label{tab:unloading_interval}
\begin{tabular}{lcccc}
\toprule
Configuration &
$t_\mathrm{hold}$ &
$t_\mathrm{unload}$ &
$E_\mathrm{ind}^{98-40}$ &
$E_\mathrm{ind}^{100-90}$ \\
&
$[\mathrm{s}]$ &
$[\mathrm{s}]$ &
$[\mathrm{GPa}]$ &
$[\mathrm{GPa}]$ \\
\midrule
No.~4 & 1  & 2  & $2.84 \pm 0.23$ & $3.00 \pm 0.25$ \\
No.~5 & 10 & 2  & $2.95 \pm 0.22$ & $2.93 \pm 0.22$ \\
No.~6 & 1  & 20 & $3.01 \pm 0.20$ & $3.69 \pm 0.31$ \\
\bottomrule
\end{tabular}
\end{table}

A pronounced dependence on the fitting interval was observed when a short holding period was combined with slow unloading. For $t_\mathrm{hold}=\SI{1}{\second}$ and $t_\mathrm{unload}=\SI{20}{\second}$, restricting the evaluation from the broader $98$--$\SI{40}{\percent}$ interval to the initial $100$--$\SI{90}{\percent}$ unloading interval increased the mean apparent indentation modulus from $\SI{3.01}{\giga\pascal}$ to $\SI{3.69}{\giga\pascal}$, corresponding to approximately $\SI{22.6}{\percent}$. Under these conditions, the brief holding phase induces substantial creep activity within the material. During the subsequent slow unloading, this persistent forward creep competes with elastic recovery. This competition artificially steepens the initial unloading slope and results in a significant overestimation of the calculated stiffness \cite{Feng2002, Jakes2024, Ngan2002}. Consequently, the local curvature immediately after the peak load does not accurately represent the overall unloading response in such cases. These results highlight the necessity of optimizing fitting windows for viscoelastic polymer systems~\cite{Sarkar2024}.

The discrepancy between fitting intervals was greatly reduced when the unloading duration was shortened to $\SI{2}{\second}$. Additionally, after a holding period of $\SI{10}{\second}$ and rapid unloading, both fitting intervals yielded nearly identical mean values of approximately $\SI{2.94}{\giga\pascal}$. These results indicate that the sensitivity of the initial unloading response decreases when the material experiences additional time-dependent deformation before unloading and when the subsequent load reduction is rapid. This confirms that sufficiently high unloading rates effectively minimize the history-dependent viscoelastic contribution to contact stiffness~\cite{Tranchida2007}.

Based on this comparison, the broad fitting interval ($\SI{40}{}$--$\SI{98}{\percent}$) was selected for the final evaluation, consistent with the established mesoscale indentation protocols for the SFRC. This interval reduces the influence of localized curvature immediately after the holding segment and provides a consistent and robust evaluation basis for all spatial measurements. Incorporating a substantially longer holding period of $\SI{300}{\second}$ in the final protocol further decreases the displacement rate at the onset of unloading to negligible levels, effectively eliminating residual viscoelastic unloading artifacts.

Comparable sensitivity was also detected during the unloading phase. Short holding periods combined with slow unloading resulted in a pronounced dependence of $E_\mathrm{ind}$ on the chosen fitting interval. This dependence was reduced when additional time-dependent deformation was allowed before unloading, allowing the creep rate to decrease, and when the subsequent unloading segment was shortened~\cite{Hardiman2016, Hochstetter1999, Ngan2002}.

\subsection{Influence of loading time at high load}
\label{subsec:loading_time}

The final assessment focuses on the loading time. Since the loading time is directly related to viscoelastic behavior, here the identical load level from the spatially resolved measurements of $P_\mathrm{max}=\SI{1750}{\milli\newton}$ with an extended holding period of $\SI{300}{\second}$ is used. This high-load condition and extended holding period of $\SI{300}{\second}$ are consistent with established microscale testing protocols for characterizing process-induced local viscoelastic fluctuations in SFRC. The high load was selected to ensure that the resulting displacement produced a pronounced and stable time-dependent signal during the holding segment. Thus, the final force selection represents a compromise among displacement resolution, spatial sampling scale, and the geometrical constraints imposed by increasing indentation size, allowing the indentation volume to act as a representative mesoscale sampling element~\cite{Hardiman2012}.

\subsubsection{High load assessment}

An independently obtained high-load dataset provides an additional reference for the sensitivity of the apparent indentation modulus to the indentation force. At $P_\mathrm{max}=\SI{1750}{\milli\newton}$, the final long-hold protocol yielded an apparent indentation modulus of $2.91\pm\SI{0.20}{\giga\pascal}$ and a maximum indentation depth of $24.16\pm\SI{0.91}{\micro\metre}$. The apparent indentation modulus therefore remained within the range observed for the controlled measurements $\SI{500}{\milli\newton}$ and $\SI{1000}{\milli\newton}$, despite the substantially larger indentation depth, indicating the absence of pronounced effects of indentation size in this depth regime \cite{Oliver1992, Tranchida2007}.

However, the $\SI{1750}{\milli\newton}$ condition was conducted with a substantially longer holding period of $\SI{300}{\second}$ and a different unloading duration. Therefore, it does not represent a third level of the controlled load comparison. Instead, this result serves as an independent consistency reference, indicating that the apparent indentation modulus remains of the same order throughout the extended load range investigated in this study. The higher load was subsequently retained in the final protocol because the associated increase in indentation depth provides a sufficiently pronounced displacement signal for characterizing the time-dependent indentation response.

\subsubsection{Loading time assessment}

The effect of the loading ramp was examined independently at $P_\mathrm{max}=\SI{1750}{\milli\newton}$. Loading durations of $\SI{1}{\second}$, $\SI{3}{\second}$, $\SI{6}{\second}$, and $\SI{10}{\second}$ were assessed, with holding and unloading durations fixed at $\SI{300}{\second}$ and $\SI{4}{\second}$, respectively. For each loading duration, $40$ indentation positions were analyzed.

Table~\ref{tab:loading_time_comparison} presents the apparent indentation modulus $E_\mathrm{ind}$, the absolute indentation creep depth increment $\Delta h_\mathrm{CR}$, and the relative indentation creep $C_\mathrm{IT}$ for the four investigated loading durations. The relative indentation creep, determined according to Eq.~\eqref{eq:relative_indentation_creep}, serves as an additional measure of the time-dependent response by normalizing the holding-period displacement increment with respect to the indentation depth at the beginning of the holding period. All quantities are reported as mean values with corresponding standard deviations.

\begin{table}[!h]
\centering
\small
\caption{Influence of loading duration on the apparent indentation modulus, absolute indentation creep depth increment, and relative indentation creep of PBT~GF0 at $P_\mathrm{max}=\SI{1750}{\milli\newton}$, $t_\mathrm{hold}=\SI{300}{\second}$, and $t_\mathrm{unload}=\SI{4}{\second}$. Each configuration comprises $40$ indentation positions. Values are reported as mean $\pm$ one standard deviation.}
\label{tab:loading_time_comparison}
\begin{tabular}{lcccc}
\toprule
Config. &
$t_\mathrm{load}$ &
$E_\mathrm{ind}$ &
$\Delta h_\mathrm{CR}$ &
$C_\mathrm{IT}$ \\
&
$[\mathrm{s}]$ &
$[\mathrm{GPa}]$ &
$[\mu\mathrm{m}]$ &
$[\%]$ \\
\midrule
No.~8  & 1  & $2.89 \pm 0.12$ & $2.88 \pm 0.13$ & $13.44 \pm 0.44$ \\
No.~9  & 3  & $3.02 \pm 0.12$ & $2.23 \pm 0.08$ & $10.49 \pm 0.42$ \\
No.~10 & 6  & $2.91 \pm 0.09$ & $2.38 \pm 0.08$ & $10.91 \pm 0.25$ \\
No.~11 & 10 & $3.03 \pm 0.07$ & $1.97 \pm 0.03$ & $9.14 \pm 0.09$ \\
\bottomrule
\end{tabular}
\end{table}

The apparent indentation modulus remained within a narrow range of $\SI{2.89}{\giga\pascal}$ to $\SI{3.03}{\giga\pascal}$, with no monotonic dependence on loading duration observed. Therefore, variation in the loading ramp influenced the stiffness-related parameter significantly less (appr. \SI{4.8}{\percent} relative to the lowest mean value) than the subsequent time-dependent displacement, confirming that the exceptionally long hold duration of $\SI{300}{\second}$ successfully allowed the viscoelastic deformation rate to diminish prior to unloading, thereby yielding drift-insensitive contact stiffness evaluations~\cite{Hardiman2016, Oliver1992}.

In contrast, both the absolute indentation creep depth increment $\Delta h_\mathrm{CR}$ and the relative indentation creep $C_\mathrm{IT}$ demonstrated a strong dependence on the preceding loading duration. The highest mean creep increment, approximately $\SI{2.88}{\micro\metre}$, was recorded after the $\SI{1}{\second}$ loading ramp, while the lowest value, approximately $\SI{1.97}{\micro\metre}$, was measured after $\SI{10}{\second}$. This represents a reduction of approximately $\SI{31.6}{\percent}$. A similar tendency was observed for $C_\mathrm{IT}$, which decreased from $\SI{13.44}{\percent}$ to $\SI{9.14}{\percent}$ between the shortest and longest loading durations, amounting to a reduction of approximately $\SI{32.0}{\percent}$. Although the intermediate values of both creep measures did not decrease monotonically, the overall trends indicate a significant influence of the preceding loading duration on the time-dependent indentation response.

In summary, varying the loading duration between $\SI{1}{\second}$ and $\SI{10}{\second}$ resulted in moderate, non-monotonic differences in $E_\mathrm{ind}$, while substantially larger variations were observed in both $\Delta h_\mathrm{CR}$ and $C_\mathrm{IT}$. This behavior can be explained by the definition $\Delta h_\mathrm{CR}=h_2-h_1$, where $h_1$ denotes the indentation depth at the beginning of the holding period~\cite{Slouf2023}. A longer loading duration permits a greater proportion of time-dependent deformation to accumulate before $P_\mathrm{max}$ is reached. Because this deformation is already included in $h_1$, it does not contribute to the subsequent holding-period increment~\cite{Cheng2004, ELSafty2012, VanLandingham2005}. The observed reduction in $C_\mathrm{IT}$ demonstrates that the relative indentation depth increment is also influenced by the preceding loading history. Therefore, meaningful comparisons of time-dependent indentation responses require identical loading and holding durations.

\subsection{Definition of the final indentation protocol}
\label{subsec:final_protocol}

Preliminary investigations demonstrate that spatially resolved indentation of PBT requires a clear distinction between parameters affecting deformation amplitude and those influencing the evaluation of the unloading response. This distinction is particularly important for viscoelastic polymers, since deformation persists beyond the maximum indentation load and continuously alters the stress and displacement fields below the indenter tip~\cite{Cheng2004, Gibson2014, Hardiman2016}.

Following the comprehensive methodological investigations, a unified indentation protocol was established for the spatial comparison of PBT~GF0 and PBT~GF30. The selected parameters are presented in Table~\ref{tab:final_indentation_protocol}.

\begin{table}[!ht]
\centering
\small
\caption{Final indentation protocol used for the spatially resolved
characterization of PBT~GF0 and PBT~GF30.}
\label{tab:final_indentation_protocol}
\begin{tabular}{lc}
\toprule
Parameter & Selected value \\
\midrule
Maximum indentation load $P_\mathrm{max}$
    & $\SI{1750}{\milli\newton}$ \\
Loading time $t_\mathrm{load}$
    & $\SI{6}{\second}$ \\
Holding time $t_\mathrm{hold}$
    & $\SI{300}{\second}$ \\
Unloading time $t_\mathrm{unload}$
    & $\SI{4}{\second}$ \\
Point-to-point spacing
    & $\SI{500}{\micro\metre}$ \\
Nominal boundary offset
    & $\SI{250}{\micro\metre}$ \\
Number of indentation positions
    & 160 \\
Unloading fitting interval
    & $98$--$\SI{40}{\percent}$ of $P_\mathrm{max}$ \\
Surface condition
    & polished \\
\bottomrule
\end{tabular}
\end{table}

Each parameter was chosen to address specific sources of experimental uncertainty. The selected spacing exceeds the experimentally determined minimum distance of approximately $\SI{400}{\micro\metre}$, and the polished surface minimizes disturbances caused by surface topography. The loading time ensures a reproducible transition to peak load without unduly prolonging the period during which time-dependent deformation may accumulate prior to the holding segment.

The high-load condition and extended holding period of $\SI{300}{\second}$ are consistent with established microscale testing protocols for characterizing process-induced local viscoelastic fluctuations in SFRC. The high load was selected to ensure that the resulting displacement produced a pronounced and stable time-dependent signal during the holding segment. Thus, the final force selection represents a compromise among displacement resolution, spatial sampling scale, and the geometrical constraints imposed by increasing indentation size, allowing the indentation volume to act as a representative mesoscale sampling element \cite{Hardiman2012}.

An extended holding period was implemented to generate a pronounced time-dependent displacement signal and to decrease the displacement rate prior to unloading. The subsequent $\SI{4}{\second}$ unloading segment restricts the duration in which creep and recovery may affect the stiffness evaluation. The preliminary comparison supports limiting the duration of unloading but does not directly establish $\SI{4}{\second}$ as an optimal value under the final testing conditions. Accordingly, this duration is treated as a fixed protocol parameter rather than an independently optimized setting. Additionally, the ($40$--$\SI{98}{\percent}$) fitting interval further reduces sensitivity to local curvature immediately after the transition from holding to unloading.

The loading-time investigation does not support the existence of a universal ramp duration beyond which indentation creep becomes independent of loading history. Instead, an identical loading ramp is necessary when comparing absolute creep increments between material systems. A loading duration of $\SI{6}{\second}$ was chosen for the final protocol due to its minimal force overshoot, relatively low response scatter, and sufficiently rapid transition into the prolonged holding segment, representing a robust balance between machine dynamics and minimizing early viscoelastic losses \cite{Slouf2023}.

Therefore, the final protocol, which incorporates a $\SI{300}{\second}$ holding period, $\SI{4}{\second}$ unloading duration, and a $\SI{98}{}$--$\SI{40}{\percent}$ fitting interval, establishes a consistent evaluation procedure and effectively reduces sensitivity to transient viscoelastic unloading artifacts near the hold-to-unload transition. This approach prevents the occurrence of "nose" anomalies and artificial stiffness inflation \cite{Feng2002, Jakes2011, Ngan2002}.

Even though the resulting indentation depth observed for a selected holding period of $\SI{300}{\second}$ in combination with a larger load is significantly deeper than for a holding time of $\SI{10}{\second}$, the resulting indentation depth still meets the reported empirical recommendations as multiples of the maximum indentation depth as documented in~\cite{Hintsala2018, Jiang2020, SudharshanPhani2019, Rossi2023}. Therefore, the same measurement  grid of 160 points with a point-to-point distance of \SI{500}{\micro\metre} is used for both PBT~GF0 and PBT~GF30 to ensure comparable measurement conditions. Additionally,  the selected spacing should be regarded as an experimentally motivated criterion rather than a universally verified, interaction-free limit for all subsequent measurements. 

The protocol, specimen preparation, measurement grid, and evaluation procedure were consistently applied to both material systems. Consequently, systematic differences in the resulting spatial distributions can be attributed primarily to variations in local material response rather than to alterations in the experimental procedure.

\section{Spatially resolved indentation response of PBT~GF0 and PBT~GF30}
\label{sec:application}

\subsection{Procedure of the spatially resolved indentation measurement}
\label{subsec:procedure_spatially_resolved}

After these preliminary investigations, spatially resolved measurements were performed using the final indentation protocol as given in Table~\ref{tab:final_indentation_protocol} that aims to minimize viscoelastic recovery artifacts during the subsequent unloading stage, which would otherwise distort the indentation modulus evaluation~\cite{Cheng2004, Hardiman2016}.

Following established mesoscale mapping strategies for SFRC~\cite{Rauter_habil2023}, a regular spatial grid containing $160$ indentation positions was applied in each investigated cross-section, as shown in Figure~\ref{fig:indentation_grid}. The complete force--displacement response was continuously recorded during loading, holding, and unloading. For the final evaluation, the unloading curve was approximated over the range from approximately $\SI{98}{\percent}$ to $\SI{40}{\percent}$ of $P_\mathrm{max}$, and the same evaluation settings were applied to all material systems. The resulting data were used to characterize the spatial indentation response and its statistical properties as summarized in Table~\ref{tab:spatial_resovled_investigations}.

\begin{table*}[t]
    \centering
    \caption{Objectives and evaluation approaches of the spatially resolved indentation investigations conducted on PBT~GF0 and PBT~GF30 at the mesoscale.}
    \label{tab:spatial_resovled_investigations}
    \small
    \begin{tabular*}{\textwidth}{    
    @{\extracolsep{\fill}}
    *{4}{>{\raggedright\arraybackslash}
         p{\dimexpr(\textwidth-8\tabcolsep)/4\relax}}
    @{}}
        \toprule
         \textbf{Methodological aspect}
        & \textbf{Objective}
        & \textbf{Experimental or analytical approach} 
        & \textbf{Resulting information} \\
        \midrule
        Spatial material response
        & Compare the local indentation response of unreinforced and reinforced PBT
        & Application of the predefined protocol to polished PBT~GF0 and PBT~GF30 cross-sections
        & Obtain spatially distributed stiffness-related and creep-related response quantities \\ \addlinespace

        Statistical description
        & Distinguish differences in mean response from differences in relative scatter
        & Evaluation of means, standard deviations, and CoV
        & Quantify the observed differences between the final spatial datasets \\
        \bottomrule
    \end{tabular*}
\end{table*}

\subsection{Overall load--displacement response}
\label{subsec:load_displacement_response}

\begin{figure}[!h]
    \centering
    \textbf{(a) PBT~GF0}\par
    \includegraphics[width=0.95\linewidth]{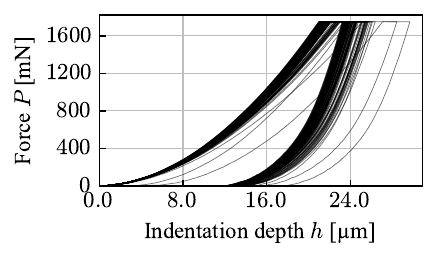}

    \vspace{1.0ex}

    \textbf{(b) PBT~GF30}\par
    \includegraphics[width=0.95\linewidth]{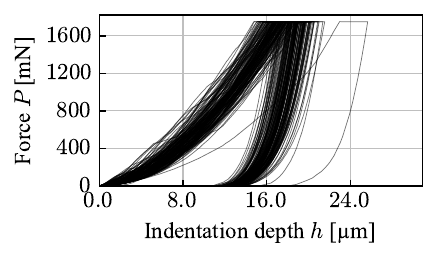}

    \caption{Experimental force--displacement responses for (a) PBT~GF0 and (b) PBT~GF30 were obtained using the final indentation protocol with $P_\mathrm{max}=\SI{1750}{\milli\newton}$, $t_\mathrm{load}=\SI{6}{\second}$, $t_\mathrm{hold}=\SI{300}{\second}$, and $t_\mathrm{unload}=\SI{4}{\second}$. The superimposed curves demonstrate the variation in indentation response across the investigated measurement positions.}
    \label{fig:representative_load_displacement}
\end{figure}

The final indentation protocol yielded distinct experimental force--displacement responses for the unreinforced and reinforced material systems. Figure~\ref{fig:representative_load_displacement} presents the superimposed force--displacement curves obtained at the investigated indentation positions for PBT~GF0 and PBT~GF30. Each curve comprises the loading segment, the additional penetration during the constant-load holding period, and the subsequent unloading response. The comparison illustrates differences in the overall indentation response and its local variability between the two material systems.

PBT~GF0 exhibited a mean maximum indentation depth of approximately $\SI{24.16}{\micro\metre}$ with a standard deviation of $\SI{0.91}{\micro\metre}$. In contrast, PBT~GF30 demonstrated a lower mean maximum depth of approximately $\SI{18.93}{\micro\metre}$ and a standard deviation of $\SI{0.94}{\micro\metre}$. This reduced penetration in the reinforced material aligns with its increased local resistance to indentation.

A time-dependent increase in indentation depth occurred during the $\SI{300}{\second}$ holding segment for both materials. However, the magnitude and spatial variability of this increase differed substantially between PBT~GF0 and PBT~GF30, as discussed in the following sections.

\subsection{Spatial distribution of the apparent indentation modulus}
\label{subsec:spatial_modulus}

\begin{figure*}[!h]
    \centering
    \textbf{(a) PBT~GF0}\par
    \includegraphics[width=0.55\linewidth]{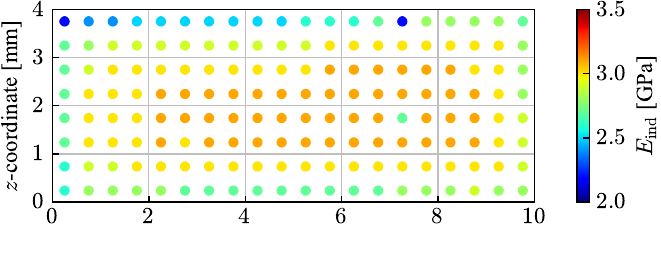}

    \vspace{1.0ex}

    \textbf{(b) PBT~GF30}\par
    \includegraphics[width=0.55\linewidth]{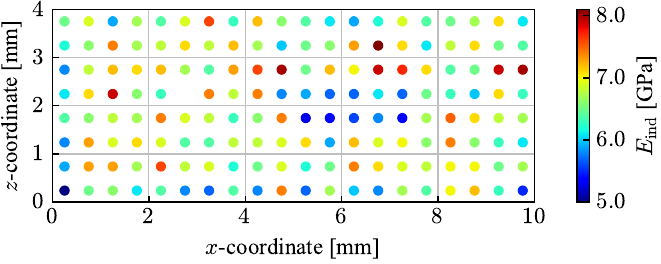}

    \caption{Spatial distribution of the apparent indentation modulus $E_\mathrm{ind}$ obtained from the unloading response of instrumented indentation measurements on the $x$--$z$ cross-sections of (a) PBT~GF0 and (b) PBT~GF30, using the final indentation protocol with $P_\mathrm{max}=\SI{1750}{\milli\newton}$ and $t_\mathrm{hold}=\SI{300}{\second}$.}
    \label{fig:spatial_distribution_indentation_modulus}
\end{figure*}

Figure~\ref{fig:spatial_distribution_indentation_modulus} shows the spatial distribution of the apparent indentation modulus across the investigated $x$--$z$ cross-sections. Retaining the coordinates of the measurement positions during evaluation enabled representation of the local response over the entire measured cross-sectional area.

PBT~GF0 exhibited a relatively narrow spatial distribution of $E_\mathrm{ind}$. Most indentation positions fell within a limited range around the mean response, which aligns with the absence of a discrete reinforcement phase. The remaining spatial scatter can be attributed to local polymer morphology, processing-related variations, and residual experimental uncertainty.

In contrast, PBT~GF30 displayed a substantially broader local modulus distribution. Regions with both low and high $E_\mathrm{ind}$ values were observed throughout the investigated cross-section. At the indentation scale, the sampled material volume may contain varying proportions and orientations of polymer matrix, glass fibers, and fiber--matrix transition regions. Therefore, the measured modulus reflects the apparent local response of the heterogeneous material volume rather than a phase-specific material constant.

The reinforced material thus exhibited both an increased mean local stiffness and a more heterogeneous spatial stiffness distribution. This finding is consistent with the process-induced microstructure of injection-molded SFRCs, for which local fiber content and fiber orientation vary throughout the cross-section~\cite{Rauter_habil2023}.

\subsection{Spatial distribution of indentation creep}
\label{subsec:spatial_creep}

\begin{figure*}[!h]
    \centering
    \textbf{(a) PBT~GF0}\par
    \includegraphics[width=0.55\linewidth]{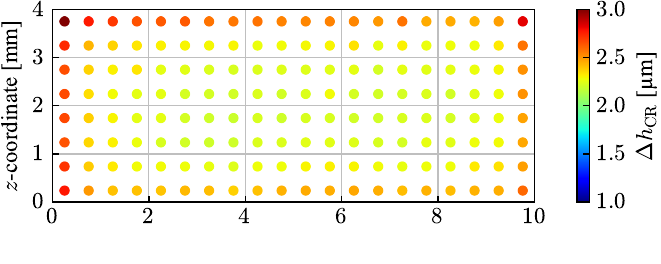}

    \vspace{1.0ex}

    \textbf{(b) PBT~GF30}\par
    \includegraphics[width=0.55\linewidth]{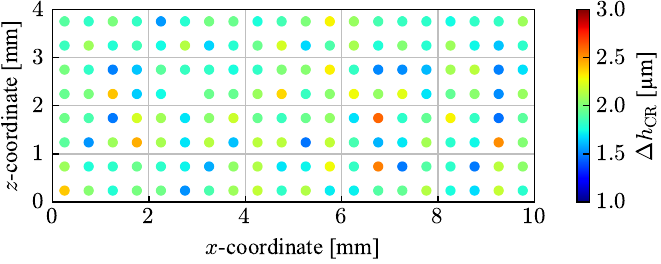}

    \caption{Spatial distribution of the absolute indentation creep depth increment $\Delta h_\mathrm{CR}$ obtained during the $\SI{300}{\second}$ holding period at $P_\mathrm{max}=\SI{1750}{\milli\newton}$ across the $x$--$z$ cross-sections of (a) PBT~GF0 and (b) PBT~GF30.}
    \label{fig:spatial_distribution_creep}
\end{figure*}

Figure~\ref{fig:spatial_distribution_creep} presents the spatial distributions of the indentation creep depth increment. The parameter $\Delta h_\mathrm{CR}$ denotes the additional penetration accumulated during the $\SI{300}{\second}$ holding period at constant $P_\mathrm{max}$.

PBT~GF0 exhibited a comparatively uniform spatial response. Despite some local deviations, the creep increment remained concentrated within a relatively narrow range across most of the cross-section. The unreinforced polymer thus served as a reference for the inherent variability of the matrix-dominated indentation response.

In the case of PBT~GF30, the mean creep increment was lower, but the local scatter was considerably greater. Distinct regions with relatively small and large $\Delta h_\mathrm{CR}$ were observed within the same cross-section. Therefore, short-fiber reinforcement did not uniformly reduce the time-dependent indentation response. Instead, the magnitude of the local deformation increment depended on the heterogeneous mechanical constraints imposed by the surrounding microstructure.

The spatial variation in creep response complements the corresponding modulus distribution. These two maps demonstrate that reinforcement-induced heterogeneity influences both the stiffness-related unloading response and the time-dependent displacement accumulated under sustained indentation load.

\subsection{Statistical comparison of PBT~GF0 and PBT~GF30}
\label{subsec:statistical_comparison}

Quantitative statistics for the spatial distributions are presented in Tables~\ref{tab:modulus_statistics}, \ref{tab:creep_statistics}, and \ref{tab:relative_creep_statistics}. All $160$ indentation positions were retained for PBT~GF0. For PBT~GF30, one position was excluded due to unstable initial contact detection, which resulted in an invalid force--displacement response, leaving $159$ valid measurements.

\begin{table}[!ht]
\centering
\small
\caption{Statistical summary of the apparent indentation modulus
$E_\mathrm{ind}$ obtained for PBT~GF0 and PBT~GF30.}
\label{tab:modulus_statistics}
\begin{tabular}{lccccc}
\toprule
Material &
\shortstack{Valid\\points} &
Mean &
\shortstack{Standard\\deviation} &
CoV &
Range \\
&
$[-]$ &
$[\mathrm{GPa}]$ &
$[\mathrm{GPa}]$ &
$[\%]$ &
$[\mathrm{GPa}]$ \\
\midrule
PBT~GF0  & 160 & 2.91 & 0.197 & 6.77 & 2.2--3.1 \\
PBT~GF30 & 159 & 6.65 & 0.597 & 8.97 & 5.0--8.1 \\
\bottomrule
\end{tabular}
\end{table}

The mean apparent indentation modulus increased from $\SI{2.91}{\giga\pascal}$ for PBT~GF0 to $\SI{6.65}{\giga\pascal}$ for PBT~GF30. Thus, incorporation of short glass fibers increased the mean local stiffness response by approximately $\SI{129}{\percent}$ under the selected indentation conditions.

The absolute scatter also increased, as indicated by the standard deviation rising from $\SI{0.197}{\giga\pascal}$ to $\SI{0.597}{\giga\pascal}$. Given the substantial difference in mean modulus between the two materials, the CoV offers a more suitable metric for comparing relative variability. The CoV increased from $6.77\,\%$ for PBT~GF0 to $\SI{8.97}{\percent}$ for PBT~GF30, representing a relative scatter increase of approximately $\SI{32.5}{\percent}$.

\begin{table}[!ht]
\centering
\small
\caption{Statistical summary of the indentation creep depth increment
$\Delta h_\mathrm{CR}$ obtained for PBT~GF0 and PBT~GF30.}
\label{tab:creep_statistics}
\begin{tabular}{lccccc}
\toprule
Material &
\shortstack{Valid\\points} &
Mean &
\shortstack{Standard\\deviation} &
CoV &
Range \\
&
$[-]$ &
$[\SI{}{\micro\metre}]$ &
$[\SI{}{\micro\metre}]$ &
$[\%]$ &
$[\SI{}{\micro\metre}]$ \\
\midrule
PBT~GF0  & 160 & 2.36 & 0.161 & 6.83 & 2.19--3.22 \\
PBT~GF30 & 159 & 1.91 & 0.233 & 12.17 & 1.47--2.60 \\
\bottomrule
\end{tabular}
\end{table}

The mean indentation creep depth showed the opposite trend. PBT~GF0 yielded a mean value of $\SI{2.36}{\micro\metre}$, while PBT~GF30 exhibited a lower mean value of $\SI{1.91}{\micro\metre}$. Consequently, the mean displacement accumulated during the holding period decreased by approximately $\SI{19}{\percent}$ after reinforcement.

Despite this reduction in the mean response, the spatial variability increased substantially. The standard deviation increased from $\SI{0.161}{\micro\metre}$ for PBT~GF0 to $\SI{0.233}{\micro\metre}$ for PBT~GF30, while the CoV increased from $\SI{6.83}{\percent}$ to $\SI{12.17}{\percent}$. The corresponding relative increase in the CoV was approximately $\SI{78}{\percent}$.

In addition to the absolute indentation creep depth increment, the relative indentation creep $C_\mathrm{IT}$ was evaluated using Equation~\eqref{eq:relative_indentation_creep}. Table~\ref{tab:relative_creep_statistics} presents the corresponding statistical results.

\begin{table}[!ht]
\centering
\small
\caption{Statistical summary of the relative indentation creep $C_\mathrm{IT}$ obtained for PBT~GF0 and PBT~GF30.}
\label{tab:relative_creep_statistics}
\begin{tabular}{lccccc}
\toprule
Material &
\shortstack{Valid\\points} &
Mean &
\shortstack{Standard\\deviation} &
CoV &
Range \\
&
$[-]$ &
$[\SI{}{\percent}]$ &
$[\SI{}{\percent}]$ &
$[\SI{}{\percent}]$ &
$[\SI{}{\percent}]$ \\
\midrule
PBT~GF0  & 160 & 10.84 & 0.458 & 4.23 & 9.4--12.8 \\
PBT~GF30 & 159 & 11.30 & 1.518 & 13.43 & 8.3--16.6 \\
\bottomrule
\end{tabular}
\end{table}

The mean relative indentation creep $C_\mathrm{IT}$ increased slightly from $\SI{10.84}{\percent}$ for PBT~GF0 to $\SI{11.29}{\percent}$ for PBT~GF30, representing an increase of approximately $\SI{4.2}{\percent}$ relative to the unreinforced material. In contrast, spatial variability increased markedly. The standard deviation increased from $0.458$ to $1.515$ percentage points, and the corresponding CoV increased from $\SI{4.23}{\percent}$ to $\SI{13.42}{\percent}$. These results suggest that short-fiber reinforcement reduced the mean absolute indentation creep depth increment but did not decrease the mean relative indentation creep under the tested loading conditions. Moreover, the increased CoV indicates substantially greater relative variability in the normalized time-dependent indentation response of PBT~GF30.

The combined statistical results indicate that short-fiber reinforcement increases the mean apparent indentation modulus and reduces the mean absolute indentation creep depth increment. In contrast, the mean relative indentation creep $C_\mathrm{IT}$ shows a slight increase. Additionally, the relative spatial variability increases for all three investigated quantities, suggesting a more heterogeneous local indentation response in PBT~GF30.

\subsection{Integrated discussion and limitations}
\label{sec:discussion}

\subsubsection{Interpretation of the apparent indentation modulus}
\label{subsec:discussion_apparent_modulus}

Interpreting the absolute magnitude of the indentation modulus requires caution, as the local contact response fundamentally differs from that of a homogeneous uniaxial tensile state. The indentation stress field is multiaxial, the sampled material volume is limited, and both plastic and time-dependent deformation may occur throughout the loading cycle~\cite{Briscoe1998, Cheng2004}. Additionally, the heterogeneous PBT~GF30 material exhibits locally varying fiber orientations and phase fractions within the indentation-affected region.

Confocal measurements confirmed the presence of pile-up around representative polished PBT~GF0 indentation imprints. Because the conventional Oliver--Pharr procedure estimates the projected contact area from the contact depth using a sink-in-based model, uncorrected pile-up may cause the actual contact area to exceed the calculated value. As a result, the measured indentation modulus is overestimated~\cite{Bolshakov1998, Oliver2004, Tranchida2007, Yan2011}. This conclusion is based on standard linear elastic tensile tests performed on PBT~GF0 where a Young's modulus of \SI{2.67}{\giga\pascal} with a standard deviation of \SI{0.30}{\giga\pascal} was obtained. 

No pointwise pile-up correction was available for complete spatial measurement grids. Therefore, $E_\mathrm{ind}$ is interpreted as an apparent indentation modulus throughout this study and should not be considered an exact intrinsic Young's modulus of either the matrix or the composite. This distinction is especially important for PBT~GF30, where individual indentation positions sample different local fiber–matrix configurations that introduce localized structural constraints and alter plastic-zone development~\cite{Hardiman2012, Slagter2023}.

However, the absence of an explicit pile-up correction does not exclude the comparative analysis conducted in this study. PBT~GF0 and PBT~GF30 were prepared, tested, and evaluated using identical experimental procedures. Consequently, the observed differences offer a consistent measure of the reinforcement effect at the investigated indentation scale, although the absolute values remain subject to the limitations of the contact-area approximation.

\subsubsection{Effect of short-fiber reinforcement on mean local response}
\label{subsec:discussion_reinforcement}

Spatial measurements demonstrate a pronounced reinforcement effect in both stiffness-related and time-dependent indentation parameters. The mean apparent indentation modulus increased from $\SI{2.91}{\giga\pascal}$ to $\SI{6.65}{\giga\pascal}$, while the mean maximum indentation depth decreased from approximately $\SI{24.16}{\micro\metre}$ to $\SI{18.93}{\micro\metre}$. These results indicate enhanced local resistance to deformation after the incorporation of short glass fibers, consistent with the expected mesoscale stiffening of SFRC~\cite{Khanna2003, Rauter2021}.

The mean creep increment also decreased from $\SI{2.36}{\micro\metre}$ to $\SI{1.91}{\micro\metre}$. Thus, the reinforcement increased resistance to initial indentation and reduced additional penetration under sustained load. This behavior is consistent with the mechanical constraint imposed by the relatively stiff reinforcement on the deformation of the surrounding polymer matrix, which restricts the development and propagation of the plastic zone beneath the indenter~\cite{Downing2000, Hardiman2012, Lee2007}.

The absolute indentation creep depth $\Delta h_\mathrm{CR}$ is not an intrinsic creep material parameter. Its magnitude depends on indentation force, loading ramp, holding duration, initial indentation depth, indenter geometry, and local mechanical constraint. Consequently, this quantity serves as a comparative local measure of the time-dependent response under the specific experimental protocol employed in this study~\cite{Briscoe1998, Gibson2014, VanLandingham2005}.

The selected examples show that a higher apparent indentation modulus $E_\mathrm{ind}$ does not necessarily coincide with a smaller absolute holding-period depth increment $\Delta h_\mathrm{CR}$, which meets the theoretical definition of these two quantities. The apparent modulus is obtained from the unloading response, which reflects instantaneous elastic recovery. In contrast, $\Delta h_\mathrm{CR}$ is determined by the displacement accumulated during the preceding holding period, which is governed by transient viscoelastic and viscoplastic flow~\cite{Briscoe1998, Ngan2002}. These two quantities assess distinct aspects of the local material response, although both are affected by the surrounding microstructure.

\subsubsection{Reinforcement-induced spatial variability}
\label{subsec:discussion_variability}

A principal outcome of the spatial analysis is that short-fiber reinforcement influences both the mean response and its relative variability. For the apparent indentation modulus $E_\mathrm{ind}$, the CoV increased from $\SI{6.77}{\percent}$ for PBT~GF0 to $\SI{8.97}{\percent}$ for PBT~GF30. This increase of approximately $\SI{32.5}{\percent}$ demonstrates that the reinforced material exhibits a broader relative distribution of local stiffness values.

This effect is even more pronounced for the time-dependent indentation response. The CoV of $\Delta h_\mathrm{CR}$ increased from $\SI{6.83}{\percent}$ to $\SI{12.17}{\percent}$, representing an increase of approximately $\SI{78}{\percent}$ in relative scatter. Although reinforcement reduces the mean creep displacement, the magnitude of this reduction varies substantially across local material regions.

The relative indentation creep $C_\mathrm{IT}$ provides an additional perspective on the time-dependent indentation response by incorporating the indentation depth at the start of the holding period. The mean value increased slightly from $\SI{10.84}{\percent}$ for PBT~GF0 to $\SI{11.29}{\percent}$ for PBT~GF30, which represents an increase of approximately $\SI{4.2}{\percent}$. In contrast, CoV increased substantially from $\SI{4.23}{\percent}$ to $\SI{13.42}{\percent}$, corresponding to an increase of approximately $\SI{217}{\percent}$ in relative scatter. Although short-fiber reinforcement reduced the mean absolute indentation creep depth increment $\Delta h_\mathrm{CR}$, the normalized creep response did not exhibit a comparable reduction. This distinction highlights the impact of varying initial indentation depths and underscores the importance of considering both absolute and relative creep measures when comparing heterogeneous material systems.

The comparatively narrow distributions observed for PBT~GF0 serve as a reference for the combined effects of matrix heterogeneity and experimental scatter. In contrast, the increased variability in PBT~GF30 is attributable to the heterogeneous microstructure introduced by short glass fibers. Individual indentation sites may exhibit differences in local fiber fraction, fiber orientation, matrix-dominated regions, and fiber–matrix interfaces. These localized variations and the resulting boundary-interaction effects, often referred to as "fiber bias" or "fiber constraint" effects, have been shown to broaden the statistical distribution of both modulus and hardness measurements in fiber-reinforced systems \cite{Hardiman2012, Hu2015, Kumar2004}. Additionally, the injection-molding process produces systematic variations in reinforcement morphology across the specimen cross-section.

\begin{figure*}[!t]
    \centering
    \begin{minipage}[t]{0.49\textwidth}
        \centering
        \textbf{a)}\par
        \vspace{0.5em}
        \includegraphics[width=\linewidth]{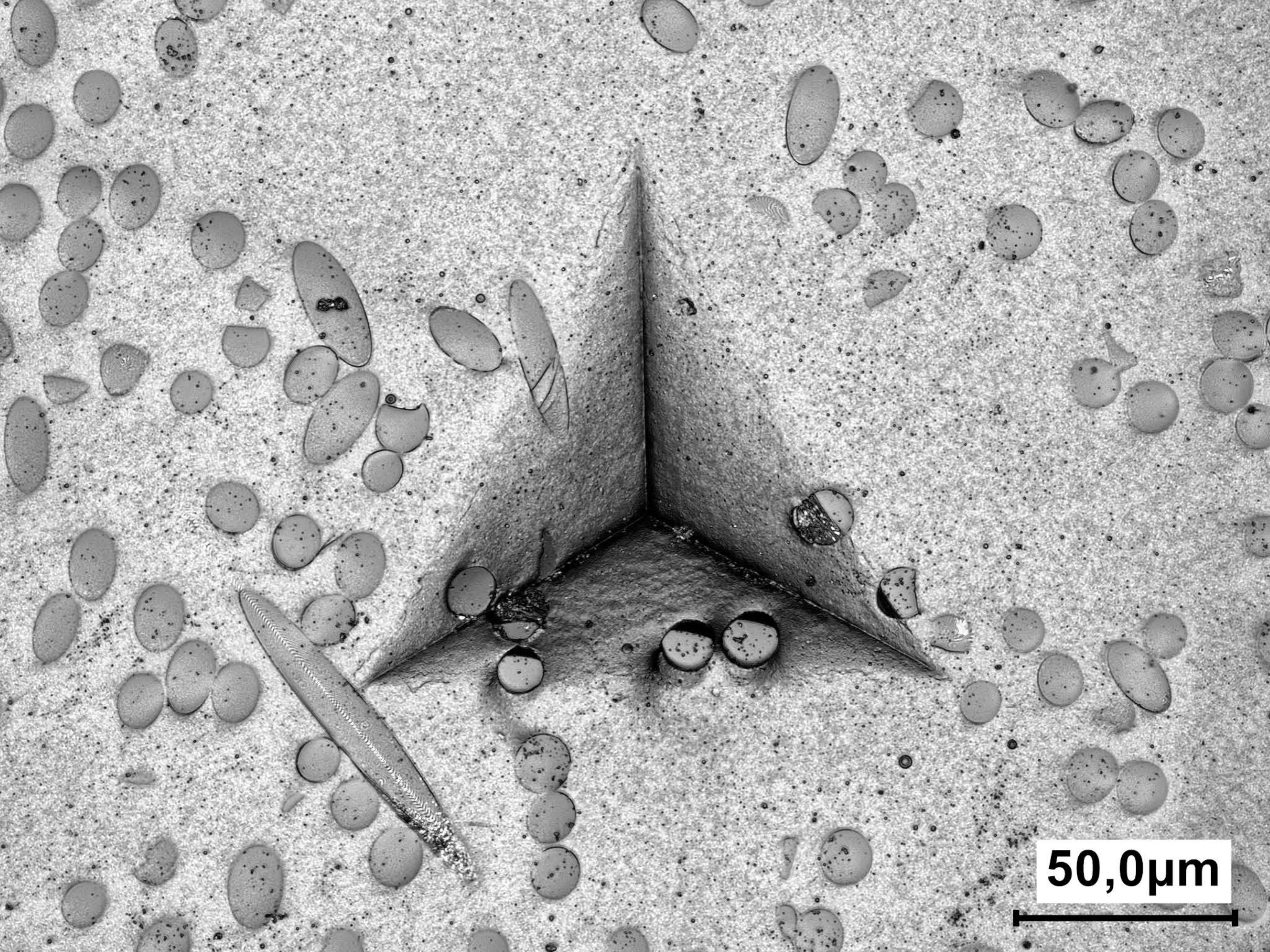}
    \end{minipage}
    \hfill
    \begin{minipage}[t]{0.49\textwidth}
        \centering
        \textbf{b)}\par
        \vspace{0.5em}
        \includegraphics[width=\linewidth]{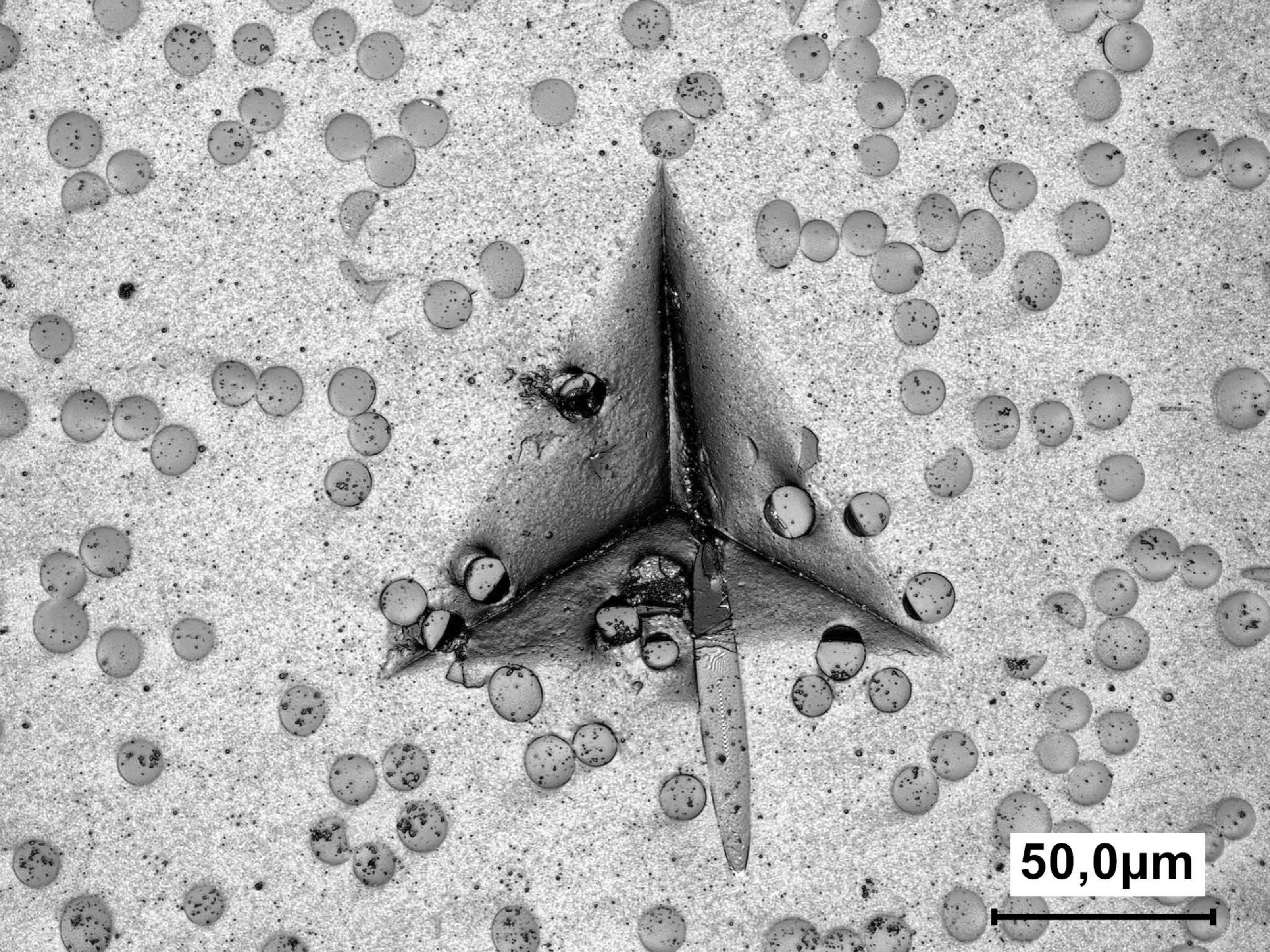}
    \end{minipage}

    \vspace{2em}

    \begin{minipage}[t]{0.49\textwidth}
        \centering
        \textbf{c)}\par
        \vspace{0.5em}
        \includegraphics[width=\linewidth]{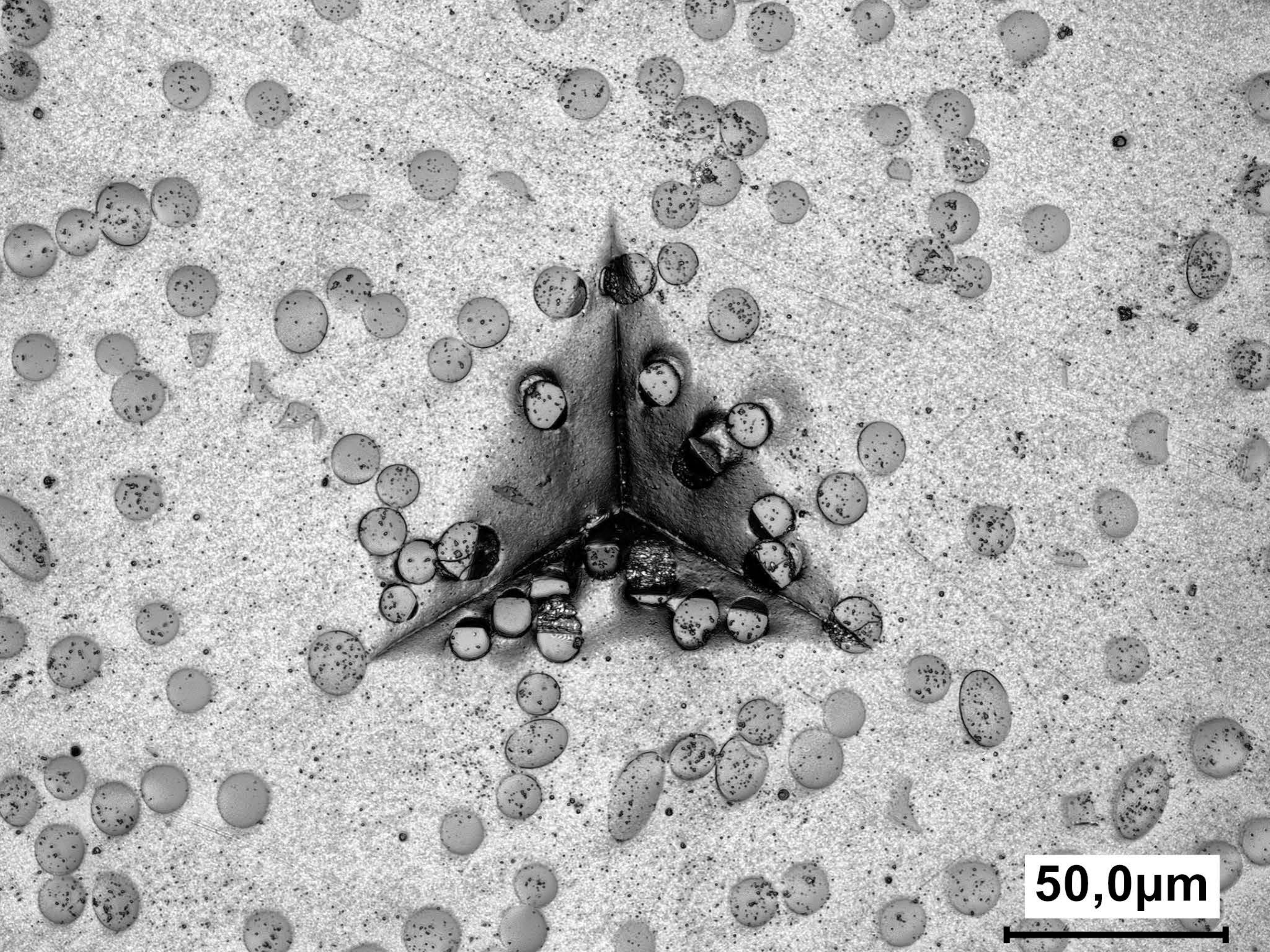}
    \end{minipage}
    \hfill
    \begin{minipage}[t]{0.49\textwidth}
        \centering
        \textbf{d)}\par
        \vspace{0.5em}
        \includegraphics[width=\linewidth]{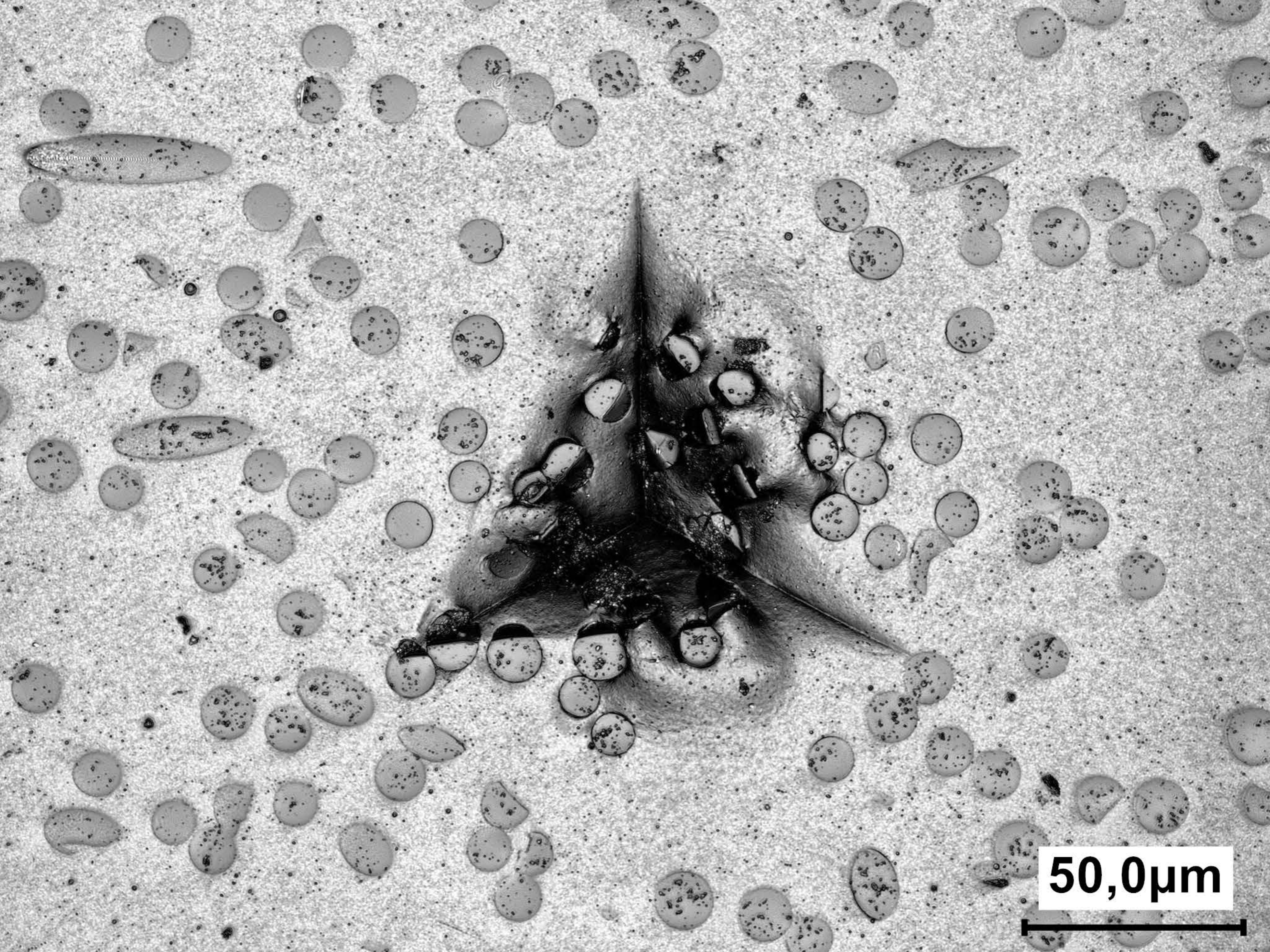}
    \end{minipage}
    \vspace{1em}
    \caption{Representative post-indentation confocal micrographs of four Berkovich imprints in PBT~GF30. Subfigures a)--d) show individual indentation sites with distinct local fiber--matrix configurations. Differences are visible in the number, spatial arrangement, and proximity of glass-fiber cross-sections relative to the residual imprints. The micrographs qualitatively illustrate variations in the local surface microstructure associated with the mesoscale material volumes involved in the indentation response. The scale bar in each micrograph corresponds to \SI{50}{\micro\metre}.}
    \label{fig:gf30_representative_indentation_imprints}
\end{figure*}

To further illustrate the microstructural heterogeneity underlying the spatially varying indentation response, Figure~\ref{fig:gf30_representative_indentation_imprints} presents representative post-indentation micrographs from four selected PBT~GF30 measurement positions. The images reveal distinct local fiber–matrix configurations, characterized by variations in the number, spatial arrangement, and proximity of visible glass-fiber cross-sections relative to the residual Berkovich imprints. These findings qualitatively demonstrate the heterogeneous material configurations encountered during indentation. The corresponding measurement coordinates and local indentation quantities are summarized in Table~\ref{tab:gf30_representative_indentation_values}, which enables identification of the selected positions in the spatial distributions shown in Figures~\ref{fig:spatial_distribution_indentation_modulus} and~\ref{fig:spatial_distribution_creep}.

\begin{table}[!h]
\centering
\small
\caption{Measurement coordinates and indentation values for the representative PBT~GF30 indentation sites are presented as shown in Figure~\ref{fig:gf30_representative_indentation_imprints}. These coordinates align with the spatial positions depicted in Figures~\ref{fig:spatial_distribution_indentation_modulus} and~\ref{fig:spatial_distribution_creep}.}
\label{tab:gf30_representative_indentation_values}
\begin{tabular}{cccccc}
\toprule
Panel &
$x$ [\SI{}{\milli\metre}] &
$z$ [\SI{}{\milli\metre}] &
$E_{\mathrm{ind}}$ [GPa] &
$\Delta h_{\mathrm{CR}}$ [\SI{}{\micro\metre}] &
$h_{\mathrm{c}}$ [\SI{}{\micro\metre}] \\
\midrule
a) & 6.25 & 1.75 & 5.6 & 1.695 & 18.561 \\
b) & 3.25 &  0.75 & 6.9 & 1.592 & 17.331 \\
c) & 6.25 &  0.75 & 7.4 & 1.828 & 15.154 \\
d) & 1.25 & 2.25 & 7.9 & 2.402 & 15.611 \\
\bottomrule
\end{tabular}
\end{table}

The corresponding apparent indentation moduli at the selected locations ranged from $\SI{5.6}{\giga\pascal}$ at position (a) to $\SI{7.9}{\giga\pascal}$ at position (d). In contrast, the absolute indentation creep depth increments varied between $\SI{1.592}{\micro\metre}$ and $\SI{2.402}{\micro\metre}$. Notably, position (d) exhibited both the highest indentation modulus and the largest creep increment among the four locations, demonstrating that these quantities are not necessarily inversely related. While the micrographs reveal differences in the visible local fiber configurations, the measured indentation response is influenced by the three-dimensional heterogeneous material volume beneath the indenter. Therefore, these observations are interpreted qualitatively and do not establish a quantitative relationship between the visible fiber distribution and the local mechanical response.

A single deterministic value does not adequately capture the full range of local indentation responses in PBT~GF30. Employing large arrays of indentations (grid indentation) with statistical descriptors enables a more comprehensive characterization of heterogeneous materials. Although statistical grid indentation was initially developed for the deconvolution and identification of discrete, homogeneous phase properties~\cite{Constantinides2006, Ulm2010}, the present study applies this spatial mapping technique to characterize the continuous statistical distribution of mesoscale apparent properties. Spatial maps and related statistical descriptors, such as the CoV and probability density distributions, provide essential insights into the magnitude and variability of the local composite response~\cite{RauterIndentation2021}. These data are especially important when local microstructural heterogeneity is incorporated into probabilistic material models.

However, the current measurements alone do not provide a comprehensive stochastic random-field characterization. Such a representation requires quantification of the spatial dependence between measurement locations, for example, through covariance or autocorrelation functions and corresponding correlation lengths. The present results establish experimental marginal statistics and spatial response maps that can serve as input for, or be supplemented by, subsequent spatial stochastic analyses.

\section{Conclusions}
\label{sec:conclusions}

A combined methodological and spatially resolved indentation approach was employed to characterize the apparent local stiffness and time-dependent indentation response of injection-molded PBT~GF0 and PBT~GF30. The methodological analysis identified experimental conditions that allow differentiation between local response variability and significant measurement-parameter effects.

The spacing analysis demonstrated that the effect of a preceding indentation became negligible at distances of approximately $\SI{400}{\micro\metre}$ under a maximum load of $\SI{1750}{\milli\newton}$. Consequently, a conservative spacing of $\SI{500}{\micro\metre}$ was selected for the final spatial grid. Additionally, polishing reduced the mean roughness depth from approximately $\SI{1.5}{\micro\metre}$–$\SI{3.0}{\micro\metre}$ for P2500-ground surfaces to approximately $\SI{0.2}{\micro\metre}$–$\SI{0.4}{\micro\metre}$, enabling reliable identification of the residual indentation geometry and surrounding pile-up.

The preliminary load comparison for PBT~GF0 demonstrated that increasing the maximum load from $\SI{500}{\milli\newton}$ to $\SI{1000}{\milli\newton}$ resulted in a substantial increase in calculated contact depth, whereas the corresponding change in apparent indentation modulus remained moderate. At $\SI{500}{\milli\newton}$, extending the holding duration from $\SI{1}{\second}$ to $\SI{20}{\second}$ led to a moderate increase in the mean apparent indentation modulus from $\SI{2.84}{\giga\pascal}$ to $\SI{2.96}{\giga\pascal}$. These findings, however, were derived from separate preliminary measurement series. The influence of unloading duration varied depending on the specific testing conditions, and the choice of unloading fitting interval substantially affected the evaluated modulus under certain circumstances.

Varying the loading duration between $\SI{1}{\second}$ and $\SI{10}{\second}$ produced apparent indentation moduli ranging from approximately $\SI{2.9}{\giga\pascal}$ to $\SI{3.0}{\giga\pascal}$, with no monotonic dependence on loading time. However, the subsequent creep depth varied substantially because different fractions of the time-dependent deformation accumulated during the preceding loading ramp. These findings indicate that the absolute indentation creep quantities are directly comparable only when an identical temporal loading protocol is applied.

Based on the methodological assessment, a final protocol consisting of $P_\mathrm{max}=\SI{1750}{\milli\newton}$, $t_\mathrm{load}=\SI{6}{\second}$, $t_\mathrm{hold}=\SI{300}{\second}$, and $t_\mathrm{unload}=\SI{4}{\second}$ was implemented for the spatial measurement grids in PBT~GF0 and PBT~GF30. The unloading response was consistently evaluated on the $\SI{98}{}$--$\SI{40}{\percent}$ interval of $P_\mathrm{max}$.

The mean apparent indentation modulus increased from $\SI{2.91}{\giga\pascal}$ for PBT~GF0 to $\SI{6.65}{\giga\pascal}$ for PBT~GF30, representing an increase of approximately $129\,\%$. In contrast, the mean creep depth decreased from $\SI{2.36}{\micro\metre}$ to $\SI{1.91}{\micro\metre}$, a reduction of approximately $19\,\%$. Thus, short-fiber reinforcement enhanced the local resistance to both initial and sustained indentation deformation.

Simultaneously, the relative spatial variability increased. The CoV of $E_\mathrm{ind}$ increased from $\SI{6.77}{\percent}$ to $\SI{8.97}{\percent}$, while the CoV of $\Delta h_\mathrm{CR}$ increased from $\SI{6.83}{\percent}$ to $\SI{12.17}{\percent}$. Therefore, reinforcement reduced the mean time-dependent indentation displacement, but resulted in a substantially more heterogeneous local response.

Confocal measurements confirmed local pile-up around representative PBT indentation imprints. As no pointwise contact-area correction was applied, the reported modulus values represent apparent indentation moduli obtained using a consistent Oliver–Pharr evaluation rather than exact intrinsic Young's moduli. Within this limitation, the established methodology offers a consistent experimental basis for quantifying both the mean local response and the spatial variability of SFRCs.

\bmsubsection*{Author Contributions}
Eduard Klatt: Conceptualization, Methodology, Software, Data Curation, Investigation, Validation, Formal analysis, Visualization, Writing - Original Draft, Writing - Review \& Editing\\
Natalie Rauter: Conceptualization, Methodology, Validation, Supervision, Project administration, Resources, Writing - Original Draft, Writing - Review \& Editing

\bmsubsection*{Conflicts of Interest}
The authors declare no conflicts of interest.

\bmsubsection*{Data Availability Statement}
The data that support the findings of this study are openly available in Zenodo at 10.5281/zenodo.23766180.

\end{document}

%% file: images/02_TensileBar_microstructure_2_paper.pdf_tex
\begingroup%
  \makeatletter%
  \providecommand\color[2][]{%
    \errmessage{(Inkscape) Color is used for the text in Inkscape, but the package 'color.sty' is not loaded}%
    \renewcommand\color[2][]{}%
  }%
  \providecommand\transparent[1]{%
    \errmessage{(Inkscape) Transparency is used (non-zero) for the text in Inkscape, but the package 'transparent.sty' is not loaded}%
    \renewcommand\transparent[1]{}%
  }%
  \providecommand\rotatebox[2]{#2}%
  \newcommand*\fsize{\dimexpr\f@size pt\relax}%
  \newcommand*\lineheight[1]{\fontsize{\fsize}{#1\fsize}\selectfont}%
  \ifx\svgwidth\undefined%
    \setlength{\unitlength}{369.25072384bp}%
    \ifx\svgscale\undefined%
      \relax%
    \else%
      \setlength{\unitlength}{\unitlength * \real{\svgscale}}%
    \fi%
  \else%
    \setlength{\unitlength}{\svgwidth}%
  \fi%
  \global\let\svgwidth\undefined%
  \global\let\svgscale\undefined%
  \makeatother%
  \begin{picture}(1,0.97178557)%
    \lineheight{1}%
    \setlength\tabcolsep{0pt}%
    \put(0,0){\includegraphics[width=\unitlength,page=1]{02_TensileBar_microstructure_2_paper.pdf}}%
    \put(0.90611306,0.41281283){\color[rgb]{0,0,0}\makebox(0,0)[lt]{\lineheight{1.25}\smash{\begin{tabular}[t]{l}$l$\end{tabular}}}}%
    \put(0.11824152,0.49541373){\color[rgb]{0,0,0}\makebox(0,0)[lt]{\lineheight{1.25}\smash{\begin{tabular}[t]{l}$x_1^{\mathrm{M}}$\end{tabular}}}}%
    \put(-0.00046212,0.63054652){\color[rgb]{0,0,0}\makebox(0,0)[lt]{\lineheight{1.25}\smash{\begin{tabular}[t]{l}$x_2^{\mathrm{M}}$\end{tabular}}}}%
    \put(0,0){\includegraphics[width=\unitlength,page=2]{02_TensileBar_microstructure_2_paper.pdf}}%
    \put(0.51425006,0.87213927){\color[rgb]{0,0,0}\makebox(0,0)[lt]{\lineheight{1.25}\smash{\begin{tabular}[t]{l}$L$\end{tabular}}}}%
    \put(0,0){\includegraphics[width=\unitlength,page=3]{02_TensileBar_microstructure_2_paper.pdf}}%
    \put(0.17511358,0.048563){\color[rgb]{0,0,0}\makebox(0,0)[lt]{\lineheight{1.25}\smash{\begin{tabular}[t]{l}$x_1^{\mathrm{m}}$\end{tabular}}}}%
    \put(0.05640968,0.17963347){\color[rgb]{0,0,0}\makebox(0,0)[lt]{\lineheight{1.25}\smash{\begin{tabular}[t]{l}$x_2^{\mathrm{m}}$\end{tabular}}}}%
    \put(0,0){\includegraphics[width=\unitlength,page=4]{02_TensileBar_microstructure_2_paper.pdf}}%
    \put(0.37822746,0.95851369){\color[rgb]{0,0,0}\makebox(0,0)[lt]{\lineheight{1.25}\smash{\begin{tabular}[t]{l}$\text{Component level}$\end{tabular}}}}%
    \put(0,0){\includegraphics[width=\unitlength,page=5]{02_TensileBar_microstructure_2_paper.pdf}}%
    \put(0.46759748,0.56447249){\color[rgb]{0,0,0}\makebox(0,0)[lt]{\lineheight{1.25}\smash{\begin{tabular}[t]{l}$\text{Microstructure}$\end{tabular}}}}%
    \put(0.36394579,0.75027086){\color[rgb]{0,0,0}\makebox(0,0)[lt]{\lineheight{1.25}\smash{\begin{tabular}[t]{l}$P(\mathbf{x}^{\mathrm{M}})$\end{tabular}}}}%
    \put(0,0){\includegraphics[width=\unitlength,page=6]{02_TensileBar_microstructure_2_paper.pdf}}%
    \put(0.00855989,0.45479092){\color[rgb]{0,0,0}\makebox(0,0)[lt]{\lineheight{1.25}\smash{\begin{tabular}[t]{l}$x_3^{\mathrm{M}}$\end{tabular}}}}%
    \put(0,0){\includegraphics[width=\unitlength,page=7]{02_TensileBar_microstructure_2_paper.pdf}}%
    \put(0.06543178,0.0079404){\color[rgb]{0,0,0}\makebox(0,0)[lt]{\lineheight{1.25}\smash{\begin{tabular}[t]{l}$x_3^{\mathrm{m}}$\end{tabular}}}}%
    \put(0.39041429,0.00387786){\color[rgb]{0,0,0}\makebox(0,0)[lt]{\lineheight{1.25}\smash{\begin{tabular}[t]{l}$E_{\text{f}}$\end{tabular}}}}%
    \put(0,0){\includegraphics[width=\unitlength,page=8]{02_TensileBar_microstructure_2_paper.pdf}}%
    \put(0.5163449,0.00387795){\color[rgb]{0,0,0}\makebox(0,0)[lt]{\lineheight{1.25}\smash{\begin{tabular}[t]{l}$E_{\text{m}}$\end{tabular}}}}%
    \put(0,0){\includegraphics[width=\unitlength,page=9]{02_TensileBar_microstructure_2_paper.pdf}}%
    \put(0.283569,0.56870296){\color[rgb]{0,0,0}\makebox(0,0)[lt]{\lineheight{1.25}\smash{\begin{tabular}[t]{l}$V$\end{tabular}}}}%
    \put(0,0){\includegraphics[width=\unitlength,page=10]{02_TensileBar_microstructure_2_paper.pdf}}%
    \put(0.97517195,0.83122782){\color[rgb]{0,0,0}\makebox(0,0)[lt]{\lineheight{1.25}\smash{\begin{tabular}[t]{l}$F$\end{tabular}}}}%
    \put(0.24446941,0.43921773){\color[rgb]{0,0,0}\makebox(0,0)[lt]{\lineheight{1.25}\smash{\begin{tabular}[t]{l}$\partial V$\end{tabular}}}}%
    \put(0,0){\includegraphics[width=\unitlength,page=11]{02_TensileBar_microstructure_2_paper.pdf}}%
    \put(0.33547878,0.30555458){\color[rgb]{0,0,0}\makebox(0,0)[lt]{\lineheight{1.25}\smash{\begin{tabular}[t]{l}$d$\end{tabular}}}}%
  \end{picture}%
\endgroup%

%% file: images/06_Zugprobe_1A_plane_xz_Nanoindent_5_paper.pdf_tex
\begingroup%
  \makeatletter%
  \providecommand\color[2][]{%
    \errmessage{(Inkscape) Color is used for the text in Inkscape, but the package 'color.sty' is not loaded}%
    \renewcommand\color[2][]{}%
  }%
  \providecommand\transparent[1]{%
    \errmessage{(Inkscape) Transparency is used (non-zero) for the text in Inkscape, but the package 'transparent.sty' is not loaded}%
    \renewcommand\transparent[1]{}%
  }%
  \providecommand\rotatebox[2]{#2}%
  \newcommand*\fsize{\dimexpr\f@size pt\relax}%
  \newcommand*\lineheight[1]{\fontsize{\fsize}{#1\fsize}\selectfont}%
  \ifx\svgwidth\undefined%
    \setlength{\unitlength}{218.4050421bp}%
    \ifx\svgscale\undefined%
      \relax%
    \else%
      \setlength{\unitlength}{\unitlength * \real{\svgscale}}%
    \fi%
  \else%
    \setlength{\unitlength}{\svgwidth}%
  \fi%
  \global\let\svgwidth\undefined%
  \global\let\svgscale\undefined%
  \makeatother%
  \begin{picture}(1,0.70459735)%
    \lineheight{1}%
    \setlength\tabcolsep{0pt}%
    \put(0,0){\includegraphics[width=\unitlength,page=1]{06_Zugprobe_1A_plane_xz_Nanoindent_5_paper.pdf}}%
    \put(0.33805983,0.37221861){\color[rgb]{0,0,0}\rotatebox{61.24587237}{\makebox(0,0)[lt]{\lineheight{1.25}\smash{\begin{tabular}[t]{l}$\text{Loading}$\end{tabular}}}}}%
    \put(0,0){\includegraphics[width=\unitlength,page=2]{06_Zugprobe_1A_plane_xz_Nanoindent_5_paper.pdf}}%
    \put(0.68453612,0.63957494){\color[rgb]{0,0,0}\makebox(0,0)[lt]{\lineheight{1.25}\smash{\begin{tabular}[t]{l}$\text{Hold}$\end{tabular}}}}%
    \put(0.66952706,0.32702922){\color[rgb]{0,0,0}\rotatebox{64.46232687}{\makebox(0,0)[lt]{\lineheight{1.25}\smash{\begin{tabular}[t]{l}$\text{Unloading}$\end{tabular}}}}}%
    \put(0,0){\includegraphics[width=\unitlength,page=3]{06_Zugprobe_1A_plane_xz_Nanoindent_5_paper.pdf}}%
    \put(-0.00066078,0.60208767){\color[rgb]{0,0,0}\makebox(0,0)[lt]{\lineheight{1.25}\smash{\begin{tabular}[t]{l}$P_{max}$\end{tabular}}}}%
    \put(0.84296381,0.02503804){\color[rgb]{0,0,0}\makebox(0,0)[lt]{\lineheight{1.25}\smash{\begin{tabular}[t]{l}$h_{2}$\end{tabular}}}}%
    \put(0.89901178,0.2063573){\color[rgb]{0,0,0}\makebox(0,0)[lt]{\lineheight{1.25}\smash{\begin{tabular}[t]{l}$S$\end{tabular}}}}%
    \put(0.49546766,0.02503804){\color[rgb]{0,0,0}\makebox(0,0)[lt]{\lineheight{1.25}\smash{\begin{tabular}[t]{l}$h_{1}$\end{tabular}}}}%
    \put(0,0){\includegraphics[width=\unitlength,page=4]{06_Zugprobe_1A_plane_xz_Nanoindent_5_paper.pdf}}%
    \put(0.9641097,0.00443417){\color[rgb]{0,0,0}\makebox(0,0)[lt]{\lineheight{1.25}\smash{\begin{tabular}[t]{l}$h$\end{tabular}}}}%
    \put(0,0){\includegraphics[width=\unitlength,page=5]{06_Zugprobe_1A_plane_xz_Nanoindent_5_paper.pdf}}%
    \put(0.60755116,0.02503804){\color[rgb]{0,0,0}\makebox(0,0)[lt]{\lineheight{1.25}\smash{\begin{tabular}[t]{l}$h_{\mathrm{f}}$\end{tabular}}}}%
    \put(0.10235874,0.68450334){\color[rgb]{0,0,0}\makebox(0,0)[lt]{\lineheight{1.25}\smash{\begin{tabular}[t]{l}$P$\end{tabular}}}}%
    \put(0.76703132,0.02503804){\color[rgb]{0,0,0}\makebox(0,0)[lt]{\lineheight{1.25}\smash{\begin{tabular}[t]{l}$h_{c}$\end{tabular}}}}%
    \put(0,0){\includegraphics[width=\unitlength,page=6]{06_Zugprobe_1A_plane_xz_Nanoindent_5_paper.pdf}}%
  \end{picture}%
\endgroup%

%% file: images/06_nanoindent-pile-up-1_paper.pdf_tex
\begingroup%
  \makeatletter%
  \providecommand\color[2][]{%
    \errmessage{(Inkscape) Color is used for the text in Inkscape, but the package 'color.sty' is not loaded}%
    \renewcommand\color[2][]{}%
  }%
  \providecommand\transparent[1]{%
    \errmessage{(Inkscape) Transparency is used (non-zero) for the text in Inkscape, but the package 'transparent.sty' is not loaded}%
    \renewcommand\transparent[1]{}%
  }%
  \providecommand\rotatebox[2]{#2}%
  \newcommand*\fsize{\dimexpr\f@size pt\relax}%
  \newcommand*\lineheight[1]{\fontsize{\fsize}{#1\fsize}\selectfont}%
  \ifx\svgwidth\undefined%
    \setlength{\unitlength}{247.11040091bp}%
    \ifx\svgscale\undefined%
      \relax%
    \else%
      \setlength{\unitlength}{\unitlength * \real{\svgscale}}%
    \fi%
  \else%
    \setlength{\unitlength}{\svgwidth}%
  \fi%
  \global\let\svgwidth\undefined%
  \global\let\svgscale\undefined%
  \makeatother%
  \begin{picture}(1,0.45068643)%
    \lineheight{1}%
    \setlength\tabcolsep{0pt}%
    \put(0,0){\includegraphics[width=\unitlength,page=1]{06_nanoindent-pile-up-1_paper.pdf}}%
    \put(0.3428104,0.04414249){\color[rgb]{0,0,0}\makebox(0,0)[lt]{\lineheight{1.25}\smash{\begin{tabular}[t]{l}$x$\end{tabular}}}}%
    \put(0.2705782,0.13031128){\color[rgb]{0,0,0}\makebox(0,0)[lt]{\lineheight{1.25}\smash{\begin{tabular}[t]{l}$y$\end{tabular}}}}%
    \put(0,0){\includegraphics[width=\unitlength,page=2]{06_nanoindent-pile-up-1_paper.pdf}}%
    \put(0.8059337,0.09170917){\color[rgb]{0,0,0}\makebox(0,0)[lt]{\lineheight{1.25}\smash{\begin{tabular}[t]{l}$h_c$\end{tabular}}}}%
    \put(0,0){\includegraphics[width=\unitlength,page=3]{06_nanoindent-pile-up-1_paper.pdf}}%
    \put(0.90887812,0.09170917){\color[rgb]{0,0,0}\makebox(0,0)[lt]{\lineheight{1.25}\smash{\begin{tabular}[t]{l}$h_{max}$\end{tabular}}}}%
    \put(0,0){\includegraphics[width=\unitlength,page=4]{06_nanoindent-pile-up-1_paper.pdf}}%
    \put(0.13679893,0.12602373){\color[rgb]{0,0,0}\makebox(0,0)[lt]{\lineheight{1.25}\smash{\begin{tabular}[t]{l}$h_c$\end{tabular}}}}%
    \put(0.8059337,0.43485506){\color[rgb]{0,0,0}\makebox(0,0)[lt]{\lineheight{1.25}\smash{\begin{tabular}[t]{l}$\text{Sink-in}$\end{tabular}}}}%
    \put(0.11964115,0.43485506){\color[rgb]{0,0,0}\makebox(0,0)[lt]{\lineheight{1.25}\smash{\begin{tabular}[t]{l}$\text{Pile-up}$\end{tabular}}}}%
    \put(0,0){\includegraphics[width=\unitlength,page=5]{06_nanoindent-pile-up-1_paper.pdf}}%
  \end{picture}%
\endgroup%

%% file: images/03_Zugprobe_1A_plane_yz_mould-injection_6_Paper.pdf_tex
\begingroup%
  \makeatletter%
  \providecommand\color[2][]{%
    \errmessage{(Inkscape) Color is used for the text in Inkscape, but the package 'color.sty' is not loaded}%
    \renewcommand\color[2][]{}%
  }%
  \providecommand\transparent[1]{%
    \errmessage{(Inkscape) Transparency is used (non-zero) for the text in Inkscape, but the package 'transparent.sty' is not loaded}%
    \renewcommand\transparent[1]{}%
  }%
  \providecommand\rotatebox[2]{#2}%
  \newcommand*\fsize{\dimexpr\f@size pt\relax}%
  \newcommand*\lineheight[1]{\fontsize{\fsize}{#1\fsize}\selectfont}%
  \ifx\svgwidth\undefined%
    \setlength{\unitlength}{758.22476634bp}%
    \ifx\svgscale\undefined%
      \relax%
    \else%
      \setlength{\unitlength}{\unitlength * \real{\svgscale}}%
    \fi%
  \else%
    \setlength{\unitlength}{\svgwidth}%
  \fi%
  \global\let\svgwidth\undefined%
  \global\let\svgscale\undefined%
  \makeatother%
  \begin{picture}(1,0.51589143)%
    \lineheight{1}%
    \setlength\tabcolsep{0pt}%
    \put(0,0){\includegraphics[width=\unitlength,page=1]{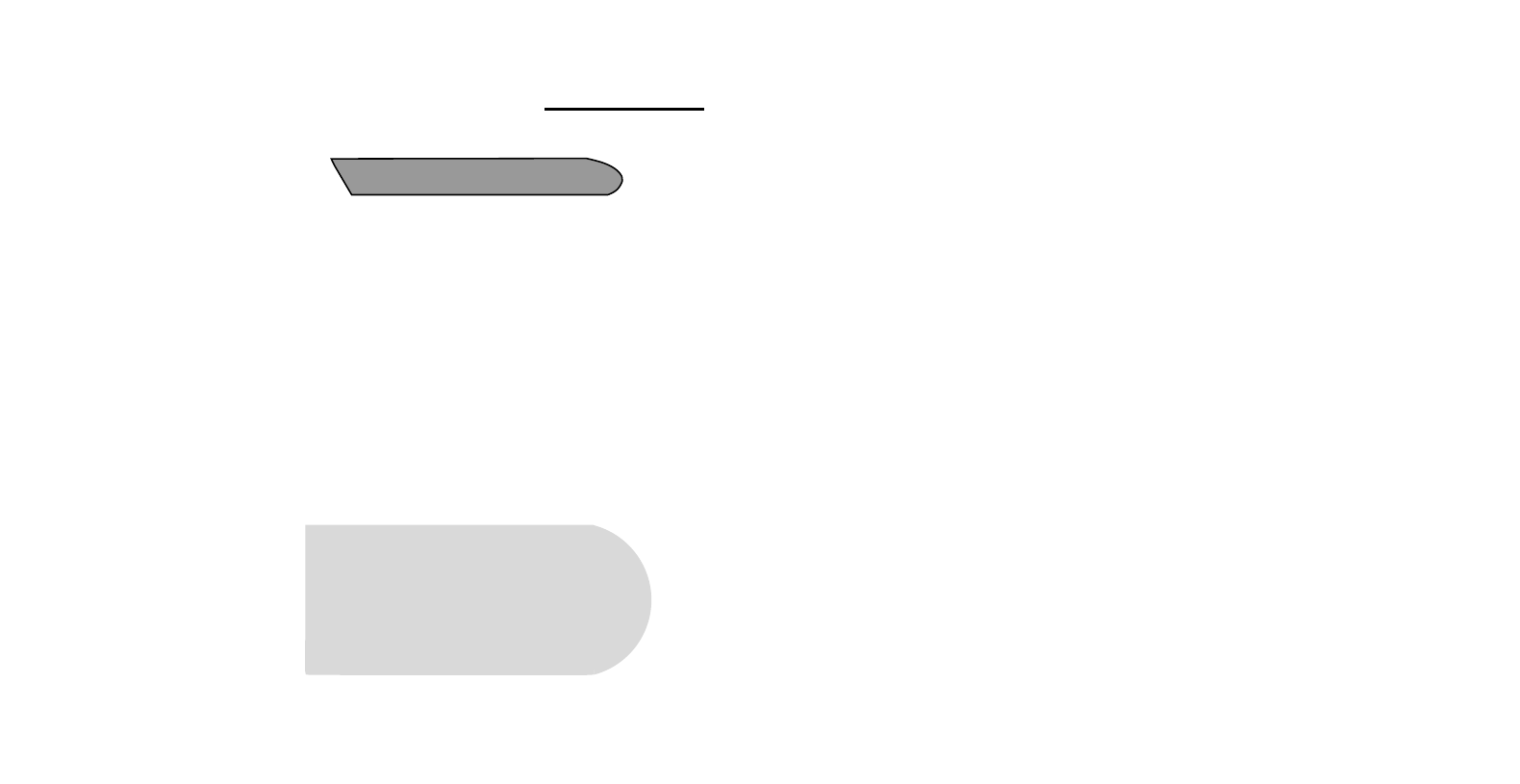}}%
    \put(0.24439856,0.5026674){\color[rgb]{0,0,0}\rotatebox{-0.87925743}{\makebox(0,0)[lt]{\lineheight{1.25}\smash{\begin{tabular}[t]{l}$\text{Melt flow direction}$ \end{tabular}}}}}%
    \put(0.38969564,0.30886923){\color[rgb]{0,0,0}\makebox(0,0)[lt]{\lineheight{1.25}\smash{\begin{tabular}[t]{l}$y$\end{tabular}}}}%
    \put(0.36049484,0.25966546){\color[rgb]{0,0,0}\makebox(0,0)[lt]{\lineheight{1.25}\smash{\begin{tabular}[t]{l}$x$\end{tabular}}}}%
    \put(0.3385954,0.31945953){\color[rgb]{0,0,0}\makebox(0,0)[lt]{\lineheight{1.25}\smash{\begin{tabular}[t]{l}$z$\end{tabular}}}}%
    \put(0,0){\includegraphics[width=\unitlength,page=2]{03_Zugprobe_1A_plane_yz_mould-injection_6_Paper.pdf}}%
    \put(0.24595219,0.01642875){\color[rgb]{0,0,0}\makebox(0,0)[lt]{\lineheight{1.25}\smash{\begin{tabular}[t]{l}$y$\end{tabular}}}}%
    \put(0.22228221,0.04255935){\color[rgb]{0,0,0}\makebox(0,0)[lt]{\lineheight{1.25}\smash{\begin{tabular}[t]{l}$z$\end{tabular}}}}%
    \put(0,0){\includegraphics[width=\unitlength,page=3]{03_Zugprobe_1A_plane_yz_mould-injection_6_Paper.pdf}}%
    \put(0.22133426,0.18934358){\color[rgb]{0,0,0}\makebox(0,0)[lt]{\lineheight{1.25}\smash{\begin{tabular}[t]{l}$v_y$\end{tabular}}}}%
    \put(0,0){\includegraphics[width=\unitlength,page=4]{03_Zugprobe_1A_plane_yz_mould-injection_6_Paper.pdf}}%
    \put(0.29943243,0.18934358){\color[rgb]{0,0,0}\makebox(0,0)[lt]{\lineheight{1.25}\smash{\begin{tabular}[t]{l}$\dot{\gamma}$\end{tabular}}}}%
    \put(0,0){\includegraphics[width=\unitlength,page=5]{03_Zugprobe_1A_plane_yz_mould-injection_6_Paper.pdf}}%
    \put(0.42486249,0.19117511){\color[rgb]{0,0,0}\makebox(0,0)[lt]{\lineheight{1.25}\smash{\begin{tabular}[t]{l}melt front\end{tabular}}}}%
    \put(0,0){\includegraphics[width=\unitlength,page=6]{03_Zugprobe_1A_plane_yz_mould-injection_6_Paper.pdf}}%
    \put(0.37392907,0.02165071){\color[rgb]{0,0,0}\makebox(0,0)[lt]{\lineheight{1.25}\smash{\begin{tabular}[t]{l}solidified melt\end{tabular}}}}%
    \put(0,0){\includegraphics[width=\unitlength,page=7]{03_Zugprobe_1A_plane_yz_mould-injection_6_Paper.pdf}}%
    \put(0.60712016,0.01665344){\color[rgb]{0,0,0}\makebox(0,0)[lt]{\lineheight{1.25}\smash{\begin{tabular}[t]{l}$y$\end{tabular}}}}%
    \put(0.58410289,0.04583966){\color[rgb]{0,0,0}\makebox(0,0)[lt]{\lineheight{1.25}\smash{\begin{tabular}[t]{l}$z$\end{tabular}}}}%
    \put(0,0){\includegraphics[width=\unitlength,page=8]{03_Zugprobe_1A_plane_yz_mould-injection_6_Paper.pdf}}%
    \put(0.09560289,0.42583604){\color[rgb]{0,0,0}\makebox(0,0)[lt]{\lineheight{1.25}\smash{\begin{tabular}[t]{l}$y$\end{tabular}}}}%
    \put(0.06067502,0.46567993){\color[rgb]{0,0,0}\makebox(0,0)[lt]{\lineheight{1.25}\smash{\begin{tabular}[t]{l}$x$\end{tabular}}}}%
    \put(0,0){\includegraphics[width=\unitlength,page=9]{03_Zugprobe_1A_plane_yz_mould-injection_6_Paper.pdf}}%
    \put(0.06485543,0.03852795){\color[rgb]{0,0,0}\makebox(0,0)[lt]{\lineheight{1.25}\smash{\begin{tabular}[t]{l}$20 \, \text{mm}$\end{tabular}}}}%
    \put(0,0){\includegraphics[width=\unitlength,page=10]{03_Zugprobe_1A_plane_yz_mould-injection_6_Paper.pdf}}%
    \put(0.11863032,0.33670557){\color[rgb]{0,0,0}\rotatebox{-90}{\makebox(0,0)[lt]{\lineheight{1.25}\smash{\begin{tabular}[t]{l}$10 \, \text{mm}$\end{tabular}}}}}%
    \put(0,0){\includegraphics[width=\unitlength,page=11]{03_Zugprobe_1A_plane_yz_mould-injection_6_Paper.pdf}}%
    \put(0.00191028,0.30307455){\color[rgb]{0,0,0}\rotatebox{-90}{\makebox(0,0)[lt]{\lineheight{1.25}\smash{\begin{tabular}[t]{l}$170 \, \text{mm} $\end{tabular}}}}}%
    \put(0,0){\includegraphics[width=\unitlength,page=12]{03_Zugprobe_1A_plane_yz_mould-injection_6_Paper.pdf}}%
    \put(0.13247844,0.23581205){\color[rgb]{0,0,0}\rotatebox{-90}{\makebox(0,0)[lt]{\lineheight{1.25}\smash{\begin{tabular}[t]{l}$\text{cross section}$\end{tabular}}}}}%
    \put(0,0){\includegraphics[width=\unitlength,page=13]{03_Zugprobe_1A_plane_yz_mould-injection_6_Paper.pdf}}%
    \put(0.08104248,0.162615){\color[rgb]{0,0,0}\rotatebox{-90}{\makebox(0,0)[lt]{\lineheight{1.25}\smash{\begin{tabular}[t]{l}$t = 4 \, \text{mm}$\end{tabular}}}}}%
    \put(0,0){\includegraphics[width=\unitlength,page=14]{03_Zugprobe_1A_plane_yz_mould-injection_6_Paper.pdf}}%
  \end{picture}%
\endgroup%

%% file: images/02_small-samples-taken-tensile-test-specimen_4_Paper.pdf_tex
\begingroup%
  \makeatletter%
  \providecommand\color[2][]{%
    \errmessage{(Inkscape) Color is used for the text in Inkscape, but the package 'color.sty' is not loaded}%
    \renewcommand\color[2][]{}%
  }%
  \providecommand\transparent[1]{%
    \errmessage{(Inkscape) Transparency is used (non-zero) for the text in Inkscape, but the package 'transparent.sty' is not loaded}%
    \renewcommand\transparent[1]{}%
  }%
  \providecommand\rotatebox[2]{#2}%
  \newcommand*\fsize{\dimexpr\f@size pt\relax}%
  \newcommand*\lineheight[1]{\fontsize{\fsize}{#1\fsize}\selectfont}%
  \ifx\svgwidth\undefined%
    \setlength{\unitlength}{229.0648203bp}%
    \ifx\svgscale\undefined%
      \relax%
    \else%
      \setlength{\unitlength}{\unitlength * \real{\svgscale}}%
    \fi%
  \else%
    \setlength{\unitlength}{\svgwidth}%
  \fi%
  \global\let\svgwidth\undefined%
  \global\let\svgscale\undefined%
  \makeatother%
  \begin{picture}(1,0.91313716)%
    \lineheight{1}%
    \setlength\tabcolsep{0pt}%
    \put(0,0){\includegraphics[width=\unitlength,page=1]{02_small-samples-taken-tensile-test-specimen_4_Paper.pdf}}%
    \put(0.04422868,0.89849334){\color[rgb]{0,0,0}\rotatebox{-0.8805768}{\makebox(0,0)[lt]{\lineheight{1.25}\smash{\begin{tabular}[t]{l}$\text{Melt flow direction}$ \end{tabular}}}}}%
    \put(0,0){\includegraphics[width=\unitlength,page=2]{02_small-samples-taken-tensile-test-specimen_4_Paper.pdf}}%
    \put(0.13743648,0.77575683){\color[rgb]{1,1,1}\makebox(0,0)[lt]{\lineheight{1.25}\smash{\begin{tabular}[t]{l}$y$\end{tabular}}}}%
    \put(0.03291031,0.72308553){\color[rgb]{1,1,1}\makebox(0,0)[lt]{\lineheight{1.25}\smash{\begin{tabular}[t]{l}$x$\end{tabular}}}}%
    \put(0,0){\includegraphics[width=\unitlength,page=3]{02_small-samples-taken-tensile-test-specimen_4_Paper.pdf}}%
    \put(0.14736851,0.06091735){\color[rgb]{1,0,0}\makebox(0,0)[lt]{\lineheight{1.25}\smash{\begin{tabular}[t]{l}$4\,\text{mm}$\end{tabular}}}}%
    \put(0,0){\includegraphics[width=\unitlength,page=4]{02_small-samples-taken-tensile-test-specimen_4_Paper.pdf}}%
    \put(0.63194702,0.06091735){\color[rgb]{1,0,0}\makebox(0,0)[lt]{\lineheight{1.25}\smash{\begin{tabular}[t]{l}$10\,\text{mm}$\end{tabular}}}}%
    \put(0,0){\includegraphics[width=\unitlength,page=5]{02_small-samples-taken-tensile-test-specimen_4_Paper.pdf}}%
  \end{picture}%
\endgroup%

%% file: images/08_ZHN_nanoindentation_2_Paper.pdf_tex
\begingroup%
  \makeatletter%
  \providecommand\color[2][]{%
    \errmessage{(Inkscape) Color is used for the text in Inkscape, but the package 'color.sty' is not loaded}%
    \renewcommand\color[2][]{}%
  }%
  \providecommand\transparent[1]{%
    \errmessage{(Inkscape) Transparency is used (non-zero) for the text in Inkscape, but the package 'transparent.sty' is not loaded}%
    \renewcommand\transparent[1]{}%
  }%
  \providecommand\rotatebox[2]{#2}%
  \newcommand*\fsize{\dimexpr\f@size pt\relax}%
  \newcommand*\lineheight[1]{\fontsize{\fsize}{#1\fsize}\selectfont}%
  \ifx\svgwidth\undefined%
    \setlength{\unitlength}{240.83530817bp}%
    \ifx\svgscale\undefined%
      \relax%
    \else%
      \setlength{\unitlength}{\unitlength * \real{\svgscale}}%
    \fi%
  \else%
    \setlength{\unitlength}{\svgwidth}%
  \fi%
  \global\let\svgwidth\undefined%
  \global\let\svgscale\undefined%
  \makeatother%
  \begin{picture}(1,1.28563504)%
    \lineheight{1}%
    \setlength\tabcolsep{0pt}%
    \put(0,0){\includegraphics[width=\unitlength,page=1]{08_ZHN_nanoindentation_2_Paper.pdf}}%
  \end{picture}%
\endgroup%

%% file: images/indentation-spacing-2_paper.pdf_tex
\begingroup%
  \makeatletter%
  \providecommand\color[2][]{%
    \errmessage{(Inkscape) Color is used for the text in Inkscape, but the package 'color.sty' is not loaded}%
    \renewcommand\color[2][]{}%
  }%
  \providecommand\transparent[1]{%
    \errmessage{(Inkscape) Transparency is used (non-zero) for the text in Inkscape, but the package 'transparent.sty' is not loaded}%
    \renewcommand\transparent[1]{}%
  }%
  \providecommand\rotatebox[2]{#2}%
  \newcommand*\fsize{\dimexpr\f@size pt\relax}%
  \newcommand*\lineheight[1]{\fontsize{\fsize}{#1\fsize}\selectfont}%
  \ifx\svgwidth\undefined%
    \setlength{\unitlength}{462.79081318bp}%
    \ifx\svgscale\undefined%
      \relax%
    \else%
      \setlength{\unitlength}{\unitlength * \real{\svgscale}}%
    \fi%
  \else%
    \setlength{\unitlength}{\svgwidth}%
  \fi%
  \global\let\svgwidth\undefined%
  \global\let\svgscale\undefined%
  \makeatother%
  \begin{picture}(1,0.73139381)%
    \lineheight{1}%
    \setlength\tabcolsep{0pt}%
    \put(0,0){\includegraphics[width=\unitlength,page=1]{indentation-spacing-2_paper.pdf}}%
    \put(0.84747756,0.28633947){\color[rgb]{0,0,0}\makebox(0,0)[lt]{\lineheight{1.25}\smash{\begin{tabular}[t]{l}$\SI{600}{\micro\metre}$\end{tabular}}}}%
    \put(0,0){\includegraphics[width=\unitlength,page=2]{indentation-spacing-2_paper.pdf}}%
    \put(0.15268017,0.28958068){\color[rgb]{0,0,0}\makebox(0,0)[lt]{\lineheight{1.25}\smash{\begin{tabular}[t]{l}$\SI{50}{\micro\metre}$\end{tabular}}}}%
    \put(0,0){\includegraphics[width=\unitlength,page=3]{indentation-spacing-2_paper.pdf}}%
    \put(0.44556801,0.38681679){\color[rgb]{0,0,0}\makebox(0,0)[lt]{\lineheight{1.25}\smash{\begin{tabular}[t]{l}$\text{Sequence 3}$\end{tabular}}}}%
    \put(0.44556801,0.25716869){\color[rgb]{0,0,0}\makebox(0,0)[lt]{\lineheight{1.25}\smash{\begin{tabular}[t]{l}$\text{Sequence 2}$\end{tabular}}}}%
    \put(0.44556801,0.12752073){\color[rgb]{0,0,0}\makebox(0,0)[lt]{\lineheight{1.25}\smash{\begin{tabular}[t]{l}$\text{Sequence 1}$\end{tabular}}}}%
    \put(0,0){\includegraphics[width=\unitlength,page=4]{indentation-spacing-2_paper.pdf}}%
    \put(0.25757821,0.71417851){\color[rgb]{0,0,0}\makebox(0,0)[lt]{\lineheight{1.25}\smash{\begin{tabular}[t]{l}$\SI{10}{\milli\metre}$\end{tabular}}}}%
    \put(0.53956287,0.72066092){\color[rgb]{0,0,0}\makebox(0,0)[lt]{\lineheight{1.25}\smash{\begin{tabular}[t]{l}$x$--$z$ $\text{cross-section}$\end{tabular}}}}%
    \put(0,0){\includegraphics[width=\unitlength,page=5]{indentation-spacing-2_paper.pdf}}%
    \put(0.06304365,0.56166764){\color[rgb]{0,0,0}\rotatebox{90}{\makebox(0,0)[lt]{\lineheight{1.25}\smash{\begin{tabular}[t]{l}$\SI{4}{\milli\metre}$\end{tabular}}}}}%
    \put(0,0){\includegraphics[width=\unitlength,page=6]{indentation-spacing-2_paper.pdf}}%
  \end{picture}%
\endgroup%

%% file: images/09_grid-pattern-nanoindentation-3-paper.pdf_tex
\begingroup%
  \makeatletter%
  \providecommand\color[2][]{%
    \errmessage{(Inkscape) Color is used for the text in Inkscape, but the package 'color.sty' is not loaded}%
    \renewcommand\color[2][]{}%
  }%
  \providecommand\transparent[1]{%
    \errmessage{(Inkscape) Transparency is used (non-zero) for the text in Inkscape, but the package 'transparent.sty' is not loaded}%
    \renewcommand\transparent[1]{}%
  }%
  \providecommand\rotatebox[2]{#2}%
  \newcommand*\fsize{\dimexpr\f@size pt\relax}%
  \newcommand*\lineheight[1]{\fontsize{\fsize}{#1\fsize}\selectfont}%
  \ifx\svgwidth\undefined%
    \setlength{\unitlength}{380.32915282bp}%
    \ifx\svgscale\undefined%
      \relax%
    \else%
      \setlength{\unitlength}{\unitlength * \real{\svgscale}}%
    \fi%
  \else%
    \setlength{\unitlength}{\svgwidth}%
  \fi%
  \global\let\svgwidth\undefined%
  \global\let\svgscale\undefined%
  \makeatother%
  \begin{picture}(1,0.53102471)%
    \lineheight{1}%
    \setlength\tabcolsep{0pt}%
    \put(0,0){\includegraphics[width=\unitlength,page=1]{09_grid-pattern-nanoindentation-3-paper.pdf}}%
    \put(0.98765598,0.24534982){\color[rgb]{0,0,0}\rotatebox{90}{\makebox(0,0)[lt]{\lineheight{1.25}\smash{\begin{tabular}[t]{l}$\SI{4}{\milli\meter}$\end{tabular}}}}}%
    \put(0,0){\includegraphics[width=\unitlength,page=2]{09_grid-pattern-nanoindentation-3-paper.pdf}}%
    \put(0.48309607,0.03900221){\color[rgb]{0,0,0}\makebox(0,0)[lt]{\lineheight{1.25}\smash{\begin{tabular}[t]{l}$\SI{10}{\milli\meter}$\end{tabular}}}}%
    \put(0,0){\includegraphics[width=\unitlength,page=3]{09_grid-pattern-nanoindentation-3-paper.pdf}}%
    \put(0.07522219,0.00356612){\color[rgb]{0,0,0}\makebox(0,0)[lt]{\lineheight{1.25}\smash{\begin{tabular}[t]{l}$x$\end{tabular}}}}%
    \put(-0.00046583,0.07966094){\color[rgb]{0,0,0}\makebox(0,0)[lt]{\lineheight{1.25}\smash{\begin{tabular}[t]{l}$z$\end{tabular}}}}%
    \put(0.9261503,0.5080674){\color[rgb]{0,0,0}\makebox(0,0)[lt]{\lineheight{1.25}\smash{\begin{tabular}[t]{l}$\SI{250}{\micro\metre}$\end{tabular}}}}%
    \put(0,0){\includegraphics[width=\unitlength,page=4]{09_grid-pattern-nanoindentation-3-paper.pdf}}%
    \put(0.65007338,0.5080674){\color[rgb]{0,0,0}\makebox(0,0)[lt]{\lineheight{1.25}\smash{\begin{tabular}[t]{l}$\SI{500}{\micro\metre}$\end{tabular}}}}%
    \put(0,0){\includegraphics[width=\unitlength,page=5]{09_grid-pattern-nanoindentation-3-paper.pdf}}%
    \put(0.00888408,0.45717502){\color[rgb]{0,0,0}\rotatebox{90}{\makebox(0,0)[lt]{\lineheight{1.25}\smash{\begin{tabular}[t]{l}$\SI{250}{\micro\metre}$\end{tabular}}}}}%
    \put(0.00888408,0.25208949){\color[rgb]{0,0,0}\rotatebox{90}{\makebox(0,0)[lt]{\lineheight{1.25}\smash{\begin{tabular}[t]{l}$\SI{500}{\micro\metre}$\end{tabular}}}}}%
    \put(0,0){\includegraphics[width=\unitlength,page=6]{09_grid-pattern-nanoindentation-3-paper.pdf}}%
  \end{picture}%
\endgroup%

%% file: images/06_nanoindent-pile-up-2_paper.pdf_tex
\begingroup%
  \makeatletter%
  \providecommand\color[2][]{%
    \errmessage{(Inkscape) Color is used for the text in Inkscape, but the package 'color.sty' is not loaded}%
    \renewcommand\color[2][]{}%
  }%
  \providecommand\transparent[1]{%
    \errmessage{(Inkscape) Transparency is used (non-zero) for the text in Inkscape, but the package 'transparent.sty' is not loaded}%
    \renewcommand\transparent[1]{}%
  }%
  \providecommand\rotatebox[2]{#2}%
  \newcommand*\fsize{\dimexpr\f@size pt\relax}%
  \newcommand*\lineheight[1]{\fontsize{\fsize}{#1\fsize}\selectfont}%
  \ifx\svgwidth\undefined%
    \setlength{\unitlength}{314.45524408bp}%
    \ifx\svgscale\undefined%
      \relax%
    \else%
      \setlength{\unitlength}{\unitlength * \real{\svgscale}}%
    \fi%
  \else%
    \setlength{\unitlength}{\svgwidth}%
  \fi%
  \global\let\svgwidth\undefined%
  \global\let\svgscale\undefined%
  \makeatother%
  \begin{picture}(1,1.39754763)%
    \lineheight{1}%
    \setlength\tabcolsep{0pt}%
    \put(0,0){\includegraphics[width=\unitlength,page=1]{06_nanoindent-pile-up-2_paper.pdf}}%
    \put(0.40819874,0.87888378){\color[rgb]{1,1,1}\rotatebox{13.053435}{\makebox(0,0)[lt]{\lineheight{1.25}\smash{\begin{tabular}[t]{l}$\SI{50}{\micro\meter}$\end{tabular}}}}}%
    \put(0.22318613,0.83599772){\color[rgb]{1,1,1}\rotatebox{13.053435}{\makebox(0,0)[lt]{\lineheight{1.25}\smash{\begin{tabular}[t]{l}$\SI{0}{\micro\meter}$\end{tabular}}}}}%
    \put(0.58367098,0.91908651){\color[rgb]{1,1,1}\rotatebox{13.053435}{\makebox(0,0)[lt]{\lineheight{1.25}\smash{\begin{tabular}[t]{l}$\SI{100}{\micro\meter}$\end{tabular}}}}}%
    \put(0.74997269,0.96082714){\color[rgb]{1,1,1}\rotatebox{13.053435}{\makebox(0,0)[lt]{\lineheight{1.25}\smash{\begin{tabular}[t]{l}$\SI{150}{\micro\meter}$\end{tabular}}}}}%
    \put(0.09201318,0.87246556){\color[rgb]{1,1,1}\rotatebox{12.22084}{\makebox(0,0)[lt]{\lineheight{1.25}\smash{\begin{tabular}[t]{l}$\SI{0}{\micro\meter}$\end{tabular}}}}}%
    \put(0.04970831,1.00354399){\color[rgb]{1,1,1}\rotatebox{12.22084}{\makebox(0,0)[lt]{\lineheight{1.25}\smash{\begin{tabular}[t]{l}$\SI{50}{\micro\meter}$\end{tabular}}}}}%
    \put(0.01767502,1.11510884){\color[rgb]{1,1,1}\rotatebox{12.22084}{\makebox(0,0)[lt]{\lineheight{1.25}\smash{\begin{tabular}[t]{l}$\SI{100}{\micro\meter}$\end{tabular}}}}}%
    \put(0,0){\includegraphics[width=\unitlength,page=2]{06_nanoindent-pile-up-2_paper.pdf}}%
    \put(0.92877299,1.04591529){\color[rgb]{0,0,0}\makebox(0,0)[lt]{\lineheight{1.25}\smash{\begin{tabular}[t]{l}$\SI{0}{}$\end{tabular}}}}%
    \put(0.92877299,1.18901985){\color[rgb]{0,0,0}\makebox(0,0)[lt]{\lineheight{1.25}\smash{\begin{tabular}[t]{l}$\SI{4.0}{}$\end{tabular}}}}%
    \put(0.92877299,1.26057212){\color[rgb]{0,0,0}\makebox(0,0)[lt]{\lineheight{1.25}\smash{\begin{tabular}[t]{l}$\SI{6.0}{}$\end{tabular}}}}%
    \put(0.92877299,1.33212447){\color[rgb]{0,0,0}\makebox(0,0)[lt]{\lineheight{1.25}\smash{\begin{tabular}[t]{l}$\SI{8.0}{}$\end{tabular}}}}%
    \put(0.9228102,1.3917518){\color[rgb]{0,0,0}\makebox(0,0)[lt]{\lineheight{1.25}\smash{\begin{tabular}[t]{l}$\SI{}{\micro\meter}$\end{tabular}}}}%
    \put(0.92877299,1.1174687){\color[rgb]{0,0,0}\makebox(0,0)[lt]{\lineheight{1.25}\smash{\begin{tabular}[t]{l}$\SI{2.0}{}$\end{tabular}}}}%
    \put(0.45397264,0.1203929){\color[rgb]{1,1,1}\rotatebox{16.940725}{\makebox(0,0)[lt]{\lineheight{1.25}\smash{\begin{tabular}[t]{l}$\SI{50}{\micro\meter}$\end{tabular}}}}}%
    \put(0.27229323,0.06506273){\color[rgb]{1,1,1}\rotatebox{16.940725}{\makebox(0,0)[lt]{\lineheight{1.25}\smash{\begin{tabular}[t]{l}$\SI{0}{\micro\meter}$\end{tabular}}}}}%
    \put(0.62631554,0.17239905){\color[rgb]{1,1,1}\rotatebox{16.940725}{\makebox(0,0)[lt]{\lineheight{1.25}\smash{\begin{tabular}[t]{l}$\SI{100}{\micro\meter}$\end{tabular}}}}}%
    \put(0.78940488,0.22531791){\color[rgb]{1,1,1}\rotatebox{16.940725}{\makebox(0,0)[lt]{\lineheight{1.25}\smash{\begin{tabular}[t]{l}$\SI{150}{\micro\meter}$\end{tabular}}}}}%
    \put(0.14448504,0.09493073){\color[rgb]{1,1,1}\rotatebox{12.22084}{\makebox(0,0)[lt]{\lineheight{1.25}\smash{\begin{tabular}[t]{l}$\SI{0}{\micro\meter}$\end{tabular}}}}}%
    \put(0.08309949,0.23077938){\color[rgb]{1,1,1}\rotatebox{12.22084}{\makebox(0,0)[lt]{\lineheight{1.25}\smash{\begin{tabular}[t]{l}$\SI{50}{\micro\meter}$\end{tabular}}}}}%
    \put(0.02721543,0.33280369){\color[rgb]{1,1,1}\rotatebox{12.22084}{\makebox(0,0)[lt]{\lineheight{1.25}\smash{\begin{tabular}[t]{l}$\SI{100}{\micro\meter}$\end{tabular}}}}}%
    \put(0,0){\includegraphics[width=\unitlength,page=3]{06_nanoindent-pile-up-2_paper.pdf}}%
    \put(0.92708649,0.3041557){\color[rgb]{0,0,0}\makebox(0,0)[lt]{\lineheight{1.25}\smash{\begin{tabular}[t]{l}$\SI{0}{}$\end{tabular}}}}%
    \put(0.92708649,0.5569746){\color[rgb]{0,0,0}\makebox(0,0)[lt]{\lineheight{1.25}\smash{\begin{tabular}[t]{l}$\SI{5.0}{}$\end{tabular}}}}%
    \put(0.92112371,0.64283718){\color[rgb]{0,0,0}\makebox(0,0)[lt]{\lineheight{1.25}\smash{\begin{tabular}[t]{l}$\SI{}{\micro\meter}$\end{tabular}}}}%
    \put(0.92708649,0.42818024){\color[rgb]{0,0,0}\makebox(0,0)[lt]{\lineheight{1.25}\smash{\begin{tabular}[t]{l}$\SI{2.5}{}$\end{tabular}}}}%
  \end{picture}%
\endgroup%

%% file: images/06_nanoindent-pile-up-3_paper.pdf_tex
\begingroup%
  \makeatletter%
  \providecommand\color[2][]{%
    \errmessage{(Inkscape) Color is used for the text in Inkscape, but the package 'color.sty' is not loaded}%
    \renewcommand\color[2][]{}%
  }%
  \providecommand\transparent[1]{%
    \errmessage{(Inkscape) Transparency is used (non-zero) for the text in Inkscape, but the package 'transparent.sty' is not loaded}%
    \renewcommand\transparent[1]{}%
  }%
  \providecommand\rotatebox[2]{#2}%
  \newcommand*\fsize{\dimexpr\f@size pt\relax}%
  \newcommand*\lineheight[1]{\fontsize{\fsize}{#1\fsize}\selectfont}%
  \ifx\svgwidth\undefined%
    \setlength{\unitlength}{354.31799305bp}%
    \ifx\svgscale\undefined%
      \relax%
    \else%
      \setlength{\unitlength}{\unitlength * \real{\svgscale}}%
    \fi%
  \else%
    \setlength{\unitlength}{\svgwidth}%
  \fi%
  \global\let\svgwidth\undefined%
  \global\let\svgscale\undefined%
  \makeatother%
  \begin{picture}(1,1.34615971)%
    \lineheight{1}%
    \setlength\tabcolsep{0pt}%
    \put(0,0){\includegraphics[width=\unitlength,page=1]{06_nanoindent-pile-up-3_paper.pdf}}%
    \put(0.36825882,0.71657693){\color[rgb]{0,0,0}\makebox(0,0)[lt]{\lineheight{1.25}\smash{\begin{tabular}[t]{l}$\SI{50}{\micro\meter}$\end{tabular}}}}%
    \put(0.16928507,0.71657693){\color[rgb]{0,0,0}\makebox(0,0)[lt]{\lineheight{1.25}\smash{\begin{tabular}[t]{l}$\SI{0}{\micro\meter}$\end{tabular}}}}%
    \put(0.56723257,0.71657693){\color[rgb]{0,0,0}\makebox(0,0)[lt]{\lineheight{1.25}\smash{\begin{tabular}[t]{l}$\SI{100}{\micro\meter}$\end{tabular}}}}%
    \put(0.75350568,0.71657693){\color[rgb]{0,0,0}\makebox(0,0)[lt]{\lineheight{1.25}\smash{\begin{tabular}[t]{l}$\SI{150}{\micro\meter}$\end{tabular}}}}%
    \put(-0.00005435,0.79701305){\color[rgb]{0,0,0}\makebox(0,0)[lt]{\lineheight{1.25}\smash{\begin{tabular}[t]{l}$\SI{0}{\micro\meter}$\end{tabular}}}}%
    \put(-0.00005438,0.89438339){\color[rgb]{0,0,0}\makebox(0,0)[lt]{\lineheight{1.25}\smash{\begin{tabular}[t]{l}$\SI{50}{\micro\meter}$\end{tabular}}}}%
    \put(-0.00005438,1.00868751){\color[rgb]{0,0,0}\makebox(0,0)[lt]{\lineheight{1.25}\smash{\begin{tabular}[t]{l}$\SI{100}{\micro\meter}$\end{tabular}}}}%
    \put(-0.00005438,1.13569209){\color[rgb]{0,0,0}\makebox(0,0)[lt]{\lineheight{1.25}\smash{\begin{tabular}[t]{l}$\SI{150}{\micro\meter}$\end{tabular}}}}%
    \put(0,0){\includegraphics[width=\unitlength,page=2]{06_nanoindent-pile-up-3_paper.pdf}}%
    \put(0.93678642,0.99598692){\color[rgb]{0,0,0}\makebox(0,0)[lt]{\lineheight{1.25}\smash{\begin{tabular}[t]{l}$\SI{0}{}$\end{tabular}}}}%
    \put(0.93678642,1.1426622){\color[rgb]{0,0,0}\makebox(0,0)[lt]{\lineheight{1.25}\smash{\begin{tabular}[t]{l}$\SI{2.5}{}$\end{tabular}}}}%
    \put(0.93678642,1.28386392){\color[rgb]{0,0,0}\makebox(0,0)[lt]{\lineheight{1.25}\smash{\begin{tabular}[t]{l}$\SI{5.0}{}$\end{tabular}}}}%
    \put(0.93149449,1.34101594){\color[rgb]{0,0,0}\makebox(0,0)[lt]{\lineheight{1.25}\smash{\begin{tabular}[t]{l}$\SI{}{\micro\meter}$\end{tabular}}}}%
    \put(0,0){\includegraphics[width=\unitlength,page=3]{06_nanoindent-pile-up-3_paper.pdf}}%
    \put(0.46265106,1.26806595){\color[rgb]{1,1,1}\makebox(0,0)[lt]{\lineheight{1.25}\smash{\begin{tabular}[t]{l}line for profile analysis \end{tabular}}}}%
    \put(0,0){\includegraphics[width=\unitlength,page=4]{06_nanoindent-pile-up-3_paper.pdf}}%
    \put(0.34595474,0.00296087){\color[rgb]{0,0,0}\makebox(0,0)[lt]{\lineheight{1.25}\smash{\begin{tabular}[t]{l}$\SI{40}{\micro\meter}$\end{tabular}}}}%
    \put(0.18084887,0.00296087){\color[rgb]{0,0,0}\makebox(0,0)[lt]{\lineheight{1.25}\smash{\begin{tabular}[t]{l}$\SI{0}{\micro\meter}$\end{tabular}}}}%
    \put(0.51529413,0.00296087){\color[rgb]{0,0,0}\makebox(0,0)[lt]{\lineheight{1.25}\smash{\begin{tabular}[t]{l}$\SI{80}{\micro\meter}$\end{tabular}}}}%
    \put(0.68463352,0.00296087){\color[rgb]{0,0,0}\makebox(0,0)[lt]{\lineheight{1.25}\smash{\begin{tabular}[t]{l}$\SI{120}{\micro\meter}$\end{tabular}}}}%
    \put(0.00727596,0.0622296){\color[rgb]{0,0,0}\makebox(0,0)[lt]{\lineheight{1.25}\smash{\begin{tabular}[t]{l}$\SI{0}{\micro\meter}$\end{tabular}}}}%
    \put(0.00727593,0.16383346){\color[rgb]{0,0,0}\makebox(0,0)[lt]{\lineheight{1.25}\smash{\begin{tabular}[t]{l}$\SI{1.0}{\micro\meter}$\end{tabular}}}}%
    \put(0.00727593,0.27813749){\color[rgb]{0,0,0}\makebox(0,0)[lt]{\lineheight{1.25}\smash{\begin{tabular}[t]{l}$\SI{2.0}{\micro\meter}$\end{tabular}}}}%
    \put(0.00727593,0.39667507){\color[rgb]{0,0,0}\makebox(0,0)[lt]{\lineheight{1.25}\smash{\begin{tabular}[t]{l}$\SI{3.0}{\micro\meter}$\end{tabular}}}}%
    \put(0.44458041,0.58408432){\color[rgb]{1,1,1}\makebox(0,0)[lt]{\lineheight{1.25}\smash{\begin{tabular}[t]{l}line for profile analysis\end{tabular}}}}%
    \put(0,0){\includegraphics[width=\unitlength,page=5]{06_nanoindent-pile-up-3_paper.pdf}}%
    \put(0.88784058,0.00296087){\color[rgb]{0,0,0}\makebox(0,0)[lt]{\lineheight{1.25}\smash{\begin{tabular}[t]{l}$\SI{165}{\micro\meter}$\end{tabular}}}}%
    \put(0,0){\includegraphics[width=\unitlength,page=6]{06_nanoindent-pile-up-3_paper.pdf}}%
    \put(0.00727593,0.51097922){\color[rgb]{0,0,0}\makebox(0,0)[lt]{\lineheight{1.25}\smash{\begin{tabular}[t]{l}$\SI{4.0}{\micro\meter}$\end{tabular}}}}%
    \put(0,0){\includegraphics[width=\unitlength,page=7]{06_nanoindent-pile-up-3_paper.pdf}}%
  \end{picture}%
\endgroup%